\documentclass [12pt,eqsecnum,amsfonts,aps]{revtex4}
\input epsf

\input epsf
\newcommand{\beq}{\begin{equation}}
\newcommand{\eeq}{\end{equation}}
\newcommand{\beqs}{\begin{eqnarray}}
\newcommand{\eeqs}{\end{eqnarray}}

\begin{document}
\baselineskip 6.0mm

\bigskip

\title{Transfer Matrices for the Partition Function of the Potts Model on
Toroidal Lattice Strips}

\bigskip

\author{Shu-Chiuan Chang}

\email{scchang@mail.ncku.edu.tw}

\affiliation{Department of Physics \\
National Cheng Kung University \\
Tainan 70101, Taiwan }

\author{Robert Shrock}

\email{robert.shrock@sunysb.edu}

\affiliation{C. N. Yang Institute for Theoretical Physics \\
State University of New York \\
Stony Brook, N. Y. 11794}

\bigskip

\begin{abstract}

We present a method for calculating transfer matrices for the $q$-state Potts
model partition functions $Z(G,q,v)$, for arbitrary $q$ and temperature
variable $v$, on strip graphs $G$ of the square (sq), triangular (tri), and
honeycomb (hc) lattices of width $L_y$ vertices and of arbitrarily great length
$L_x$ vertices, subject to toroidal and Klein bottle boundary conditions.  For
the toroidal case we express the partition function as $Z(\Lambda, L_y \times
L_x,q,v) = \sum_{d=0}^{L_y} \sum_j b_j^{(d)} (\lambda_{Z,\Lambda,L_y,d,j})^m$,
where $\Lambda$ denotes lattice type, $b_j^{(d)}$ are specified polynomials of
degree $d$ in $q$, $\lambda_{Z,\Lambda,L_y,d,j}$ are eigenvalues of the
transfer matrix $T_{Z,\Lambda,L_y,d}$ in the degree-$d$ subspace, and $m=L_x$
($L_x/2$) for $\Lambda=sq, \ tri \ (hc)$, respectively. An analogous formula is
given for Klein bottle strips.  We exhibit a method for calculating
$T_{Z,\Lambda,L_y,d}$ for arbitrary $L_y$.  In particular, we find some very
simple formulas for the determinant $det(T_{Z,\Lambda,L_y,d})$, and trace
$Tr(T_{Z,\Lambda,L_y})$.  Corresponding results are given for the equivalent
Tutte polynomials for these lattice strips and illustrative examples are
included.

\end{abstract}

\maketitle

\newpage
\pagestyle{plain}
\pagenumbering{arabic}

\section{Introduction}

The $q$-state Potts model has served as a valuable model for the study of phase
transitions and critical phenomena \cite{potts,wurev}.  On a lattice, or, more
generally, on a (connected) graph $G$, at temperature $T$, this model is
defined by the partition function
\beq
Z(G,q,v) = \sum_{ \{ \sigma_n \} } e^{-\beta {\cal H}}
\label{zfun}
\eeq%
with the (zero-field) Hamiltonian
\beq
{\cal H} = -J \sum_{\langle i j \rangle} \delta_{\sigma_i \sigma_j}
\label{ham}
\eeq
where $\sigma_i=1,...,q$ are the spin variables on each vertex (site)
$i \in G$;
$\beta = (k_BT)^{-1}$; and $\langle i j \rangle$ denotes pairs of adjacent
vertices.  The graph $G=G(V,E)$ is defined by its vertex set $V$ and its edge
set $E$; we denote the number of vertices of $G$ as $n=n(G)=|V|$ and the
number of edges of $G$ as $e(G)=|E|$.  We use the notation
\beq
K = \beta J \ , \quad v = e^K - 1
\label{kdef}
\eeq
so that the physical ranges are $v \ge 0$ for the Potts ferromagnet, and $-1
\le v \le 0$ for Potts antiferromagnet, corresponding to $0 \le T \le \infty$.
One defines the (reduced) free energy per site $f=-\beta F$, where $F$ is the
actual free energy, via $f(\{G\},q,v) = \lim_{n \to \infty} \ln [
Z(G,q,v)^{1/n}]$, where we use the symbol $\{G\}$ to denote the formal limit
$\lim_{n \to \infty}G$ for a given family of graphs.  In the present context,
this $n \to \infty$ limit corresponds to the limit of infinite length for a
strip graph of fixed width and some prescribed boundary conditions.

In this paper we shall present transfer matrices for the $q$-state Potts model
partition functions $Z(G,q,v)$, for arbitrary $q$ and temperature variable $v$,
on toroidal and Klein bottle strip graphs $G$ of the square, triangular, and
honeycomb lattices of width $L_y$ vertices and of arbitrarily great length
$L_x$ vertices.  We label the lattice type as $\Lambda$ and abbreviate the
three respective types as $sq$, $tri$, and $hc$.  Each strip involves a
longitudinal repetition of $m$ copies of a particular subgraph.  For the
square-lattice strips, this is a column of squares.  It is convenient to
represent the strip of the triangular lattice as obtained from the
corresponding strip of the square lattice with additional diagonal edges
connecting, say, the upper-left to lower-right vertices in each square.  In
both these cases, the length is $L_x=m$ vertices.  We represent the strip of
the honeycomb lattice in the form of bricks oriented horizontally.  In this
case, since there are two vertices in 1-1 correspondence with each horizontal
side of a brick, $L_x=2m$ vertices.  Summarizing for all of three lattices, the
relation between the number of vertices and the number of repeated copies is
\beq
L_x=\cases{ m & if $\Lambda=sq$ \ or \ $tri$ \cr
            2m & if $\Lambda=hc$ \cr } \ .
\label{lxm}
\eeq
For the toroidal case the partition function has the general form
\cite{a,bcc}
\beq
Z(\Lambda, L_y \times L_x,tor.,q,v) = \sum_{d=0}^{L_y} \sum_j b_j^{(d)} (\lambda_{Z,\Lambda,L_y,d,j})^m
\label{zgsumtor}
\eeq
where $\lambda_{Z,\Lambda,L_y,d,j}$, are eigenvalues of the transfer
matrices in the degree-$d$ subspace $T_{Z,\Lambda,L_y,d}$ and they are
independent of the length of the strip length $L_x$. The size of the
transfer matrices $T_{Z,\Lambda,L_y,d}$ is equal to $d! \
n_{Z,tor}(L_y,d)$ that will be discussed in the following
section. Here because the dimensions of $T_{Z,\Lambda,L_y,d}$ are the
same for all three lattices $\Lambda=sq,tri,hc$, we omit $\Lambda$ in
the notation. In contrast to cyclic and M\"obius strips, there can be
more than one coefficient, denoted as $b_j^{(d)}$, for each degree
$d$.  The $b_j^{(d)}$'s are polynomials of degree $d$ in $q$ and play
a role analogous to multiplicities of eigenvalues
$\lambda_{Z,\Lambda,L_y,d,j}$, although this identification is formal,
since $b_j^{(d)}$ may be zero or negative for the small physical
values of $q$.  We exhibit our method for calculating
$T_{Z,\Lambda,L_y,d}$ for arbitrary $L_y$, and we shall construct an
analogous formula for Klein bottle strips.  Explicit results for
arbitrary $L_y$ are given for (i) $T_{Z,\Lambda,L_y,d}$ with $d=L_y$,
(ii) the determinant $det(T_{Z,\Lambda,L_y,d})$, and (iii) the trace
$Tr(T_{Z,\Lambda,L_y})$.  The results in (i) and (iii) apply for
$\Lambda=sq,tri,hc$, while the results in (ii) apply for
$\Lambda=sq,hc$.  Corresponding results are given for the equivalent
Tutte polynomials for these lattice strips and illustrative examples
are included.  We have calculated the transfer matrices up to widths
$L_y=4$ for the square and honeycomb lattices and up to $L_y=3$ for
the triangular lattice.  Since the dimensions of these matrices
increase rapidly with strip width (e.g.,
$dim(T_{Z,\Lambda,L_y,d})=14,35,56,48$ for $L_y=4$ and $0 \le d \le
3$), it is not feasible to present many of the explicit results here;
instead, we concentrate on general methods and results that hold for
arbitrary $L_y$.

Various special cases of the Potts model partition function $Z(G,q,v)$ are of
interest.  For example, if one considers the case of antiferromagnetic
spin-spin coupling, $J < 0$ and takes the temperature to zero, so that
$K=-\infty$ and $v=-1$, then
\beq
Z(G,q,-1)=P(G,q)
\label{zp}
\eeq
where $P(G,q)$ is the chromatic polynomial (in $q$) expressing the number of
ways of coloring the vertices of the graph $G$ with $q$ colors such that no two
adjacent vertices have the same color \cite{bbook,rtrev}. The minimum number of colors necessary for such a coloring of $G$ is called the chromatic number, $\chi(G)$. The $q$-state Potts antiferromagnet (AF) \cite{potts,wurev} exhibits nonzero ground state entropy, $S_0 > 0$ (without frustration) for sufficiently large $q$ on a given lattice $\Lambda$ or, more generally, on a graph $G=(V,E)$. This is equivalent to a ground state degeneracy per site $W > 1$, since $S_0 = k_B \ln W$.
Thus
\beq
W(\{G\},q) = \lim_{n \to \infty} P(G,q)^{1/n} \ .
\label{w}
\eeq

For a strip graph $G_s$ of a lattice $\Lambda$ with given boundary conditions,
following our earlier notation \cite{a} we denote the sum of coefficients
(generalized multiplicities) $c_{Z,G,j}$ of the $\lambda_{Z,G,j}$ as
\beq
C_{Z,G}=\sum_{j=1}^{N_{Z,G,\lambda}} c_{Z,G,j}
\label{czsum}
\eeq
while for the chromatic polynomial as
\beq
C_{P,G}=\sum_{j=1}^{N_{P,G,\lambda}} c_{P,G,j} \ .
\label{cpsum}
\eeq
These sums are independent of the length $m$ of the strip. 
General results for these sums will be given below for the strips of interest.

We recall some previous related work.  The partition function $Z(G,q,v)$ for
the Potts model was calculated for arbitrary $q$ and $v$ on strips of the
lattice $\Lambda$ with toroidal or Klein bottle longitudinal boundary conditions was calculated for (i) $\Lambda=sq$, $L_y=2, 3$ in \cite{s3a}, 
and (ii) for $L_y=3$ on the square-lattice with next-nearest-neighbor
spin-spin couplings in \cite{ka3}.   Matrix methods for calculating chromatic polynomials were developed and used in \cite{bds,bm,baxter86,baxter87} and more recently in \cite{matmeth,matmeth2,matmeth3}.  Ref. \cite{sqtran, cyltran, tritran} developed transfer matrix methods for both $Z(G,q,v)$ and the special case $v=-1$ of chromatic polynomials on strips of the square and triangular lattices with free longitudinal boundary
conditions and used them to calculate the latter polynomials for a large
variety of widths.  These have been termed transfer matrices in the
Fortuin-Kasteleyn representation (see eq. (\ref{cluster}) below).  These
methods were applied to calculate the full Potts model partition function for
strips of the square and triangular lattices with free boundary conditions and
a number of widths in Refs. \cite{ts,tt}, and then were generalized for strips of the square, triangular and honeycomb lattices with cyclic and M\"obious strips \cite{zt, pt}.
Here we extend these transfer matrix methods for
calculating Potts model partition functions on toroidal and Klein bottle 
lattice strips and present general results for these lattice strips of
arbitrary width. Clearly, new exact results on the Potts
model are of value in their own right.

Let $G^\prime=(V,E^\prime)$ be a spanning subgraph of $G$, i.e. a subgraph
having the same vertex set $V$ and an edge set $E^\prime \subseteq E$. 
$Z(G,q,v)$ can be written as the sum \cite{kf,fk}
\beq
Z(G,q,v) = \sum_{G^\prime \subseteq G} q^{k(G^\prime)}v^{|E^\prime|}
\label{cluster}
\eeq
where $k(G^\prime)$ denotes the number of connected components of $G^\prime$.
Since we only consider connected graphs $G$, we have $k(G)=1$. The formula
(\ref{cluster}) enables one to generalize $q$ from ${\mathbb Z}_+$ to ${\mathbb
R}_+$ for physical ferromagnetic $v$.  More generally, eq. (\ref{cluster})
allows one to generalize both $q$ and $v$ to complex values, as is necessary
when studying zeros of the partition function in the complex $q$ and $v$
planes.  

The Potts model partition function $Z(G,q,v)$ is equivalent to an object of
considerable current interest in mathematical graph theory, the Tutte
polynomial, $T(G,x,y)$, given by \cite{tutte1,tutte2,tutte3}
\beq
T(G,x,y)=\sum_{G^\prime \subseteq G} (x-1)^{k(G^\prime)-k(G)}
(y-1)^{c(G^\prime)}
\label{tuttepol}
\eeq
where $c(G^\prime) = |E^\prime|+k(G^\prime)-|V|$ 
is the number of independent circuits in $G^\prime$.  Now let
\beq
x=1+\frac{q}{v}, \quad y=v+1
\label{xdef}
\eeq
so that
\beq
q=(x-1)(y-1) \ .  
\label{qxy}
\eeq
Then the equivalence between the Potts model partition function and the Tutte
polynomial for a graph $G$ is
\beq
Z(G,q,v)=(x-1)^{k(G)}(y-1)^{n(G)}T(G,x,y) \ .
\label{zt}
\eeq
Given this equivalence, we can express results either in Potts or Tutte form.
We will use both, since each has its own particular advantages.  The Potts
model form, involving the variables $q$ and $v$ is convenient for physical
applications, since $q$ specifies the number of states and determines the
universality class of the transition, and $v$ is the temperature variable.  The
Tutte form has the advantage that many expressions are simpler when written in
terms of the Tutte variables $x$ and $y$.  

 From (\ref{zt}), one can write the Tutte polynomial as 
\beq
T(\Lambda,L_y \times L_x,tor.,x,y) = \frac{1}{x-1} \sum_{d=0}^{L_y} 
\sum_j b_j^{(d)} (\lambda_{T,\Lambda,L_y,d,j})^m
\label{tgsumtor}
\eeq
where $m$ is given in terms of $L_x$ by eq. (\ref{lxm}) and it is convenient to
factor out a factor of $1/(x-1)$. (This factor is always cancelled, since the
Tutte polynomial is a polynomial in $x$ as well as $y$.)

 From eq. (\ref{zt}) it follows that \cite{a,hca}
\beq
\lambda_{Z,\Lambda,L_y,d,j} = v^{p L_y} \lambda_{T,\Lambda,L_y,d,j}
\label{lamzt}
\eeq
and
\beq
T_{Z,\Lambda,L_y,d} = v^{p L_y} T_{T,\Lambda,L_y,d}
\label{tztt}
\eeq
where
\beq
p = \cases{ 1 & if $\Lambda=sq$ \ or \ $tri$ \cr
            2 & if $\Lambda=hc$ \cr } \ .
\label{powerp}
\eeq
Note that the factor of $(x-1)$ in eq. (\ref{zt}) cancels the factor $1/(x-1)$
in eq. (\ref{tgsumtor}).

\section{Transfer Matrix Method}

The chromatic polynomials for strips with periodic boundary condition in the
longitudinal direction were discussed in Refs. \cite{matmeth,matmeth2,matmeth3}
in terms of a compatibility matrix, and the bases of the matrix are given by
$[X|\xi]$, where $X$ is a subset of $L_y$ vertices and $\xi$ is an injection
 from $X$ to $\{1,2,...,q\}$. We now use the same bases for the transfer matrix
of the full Potts model partition function. The degree $d=0$ subspace of the
transfer matrix (equivalently called the ``level 0'' subspace in
\cite{matmeth3}) corresponds to the empty set $X=\emptyset$, and the bases of
the transfer matrix are all of the possible non-crossing partitions of $L_y$
vertices.  (For the zero-temperature antiferromagnetic Potts model, a
non-nearest-neighbor requirement is imposed; this will be discussed further in
the next section.)  The dimension of this matrix is $n_{Z,tor}(L_y,0)=C_{L_y}$,
the Catalan number \cite{sqtran}, as in eq. (\ref{nzly0}) below.  The
eigenvalues of the transfer matrix for a cylindrical strip are a subset of the
eigenvalues of the transfer matrix $T_{Z,\Lambda,L_y,d=0}$ in this degree $d=0$
subspace for the corresponding toroidal strip.  The degree $d=1$ subspace of
the transfer matrix is given by all of the possible non-crossing partitions
with a color assignment to one vertex (with possible connections with other
vertices); i.e., $X$ contains one vertex, and the multiplicity is $q-1 \equiv
b^{(1)}$. This follows because there are $q$ possible ways of making this color
assignment, but one of these has to be subtracted, since the effect of all the
possible color assignments is equivalent to the choice of no specific color
assignment, which has been taken into account in the level 0 subspace.  In this
derivation and subsequent ones we assume that $q$ is a sufficiently large
integer to begin with, so that the multiplicities are positive-definite; we
then analytically continue them downward to apply in the region of small $q$
where the coefficients can be zero or negative.  For the next subspace we
consider all of the non-crossing partitions with two-color assignments to two
separated vertices (with possible connections with other vertices). Now the
multiplicity can be understood by the sieve formula of
\cite{matmeth,matmeth2,matmeth3}. Since the two assigned colors should be
different, there are $q(q-1)$ ways of making these assignments. This includes
the $q$ possible color assignments for each of the two vertices that have been
considered in level 1 and hence these must be subtracted. In doing this, the
no-color assignment was subtracted twice, and one of these has to be added
back. Therefore, the multiplicity is
\beq
q(q-1)-2q+1 = q^2-3q+1 \equiv b^{(2)} \ .
\label{bd2derive}
\eeq
Recall that a slice of cyclic strips is a tree graph, but a slice of 
toroidal strips is a circuit graph. 
For toroidal strips, these coefficients $b^{(d)}$ with $0 \le d \le 2$ are the
same as those for cyclic strips $c^{(d)}$ in Ref. \cite{cf} (see also
\cite{saleur}) because when the number of vertices is 1 or 2 a tree graph is
equivalent to a circuit graph modulo multiple edges. 
For toroidal strips, the multiplicity for three-color assignment is
\beq
q(q-1)(q-2)-3q(q-1)+3q-1 = q^3-6q^2+8q-1 \equiv b^{(3)} \ . 
\label{bd3derive}
\eeq
By similar reasoning, the multiplicity for four-color assignment is
\beqs
& & q(q-1)(q^2-3q+3)-4q(q-1)^2+4q(q-1)+2q^2-4q+1 \cr\cr
& = & q^4-8q^3+20q^2-15q+1 \equiv b^{(4)} \ . 
\label{bd4derive}
\eeqs
Further, we find 
\beqs
b^{(5)} & = & q^5-10q^4+35q^3-50q^2+24q-1 \cr\cr
b^{(6)} & = & q^6-12q^5+54q^4-112q^3+105q^2-35q+1 \cr\cr
b^{(7)} & = & q^7-14q^6+77q^5-210q^4+294q^3-196q^2+48q-1 \cr\cr
b^{(8)} & = & (q^2-3q+1)(q^6-13q^5+64q^4-147q^3+155q^2-60q+1) \cr\cr
b^{(9)} & = & q^9-18q^8+135q^7-546q^6+1287q^5-1782q^4+1386q^3-540q^2+80q-1 
\cr\cr
b^{(10)} & = & q^{10}-20q^9+170q^8-800q^7+2275q^6-4004q^5+4290q^4-2640q^3+
825q^2-99q+1 \cr\cr
& &
\label{bd5bd10}
\eeqs
and so forth for higher values of $d$. We show the calculation for these
multiplicities pictorially for $2 \le d \le 4$ in Fig. \ref{bdfigure}. Let us
define 
\beq
q = 2 + t + t^{-1} = 2 + 2\cos\theta = 4\cos^2(\theta/2)\ . 
\label{qtheta}
\eeq
Then, in terms of these variables, we find that $b^{(d)}$ can be
written as
\beq
b^{(0)}=1
\label{b0}
\eeq
\beq
b^{(1)} = t+1+t^{-1} = 1 + 2\cos\theta
\label{b1}
\eeq
and, for $d \ge 2$, 
\beq
b^{(d)} = (-1)^d (t+1+t^{-1})+t^d+t^{-d} = 
          (-1)^d (1+2\cos\theta )+2\cos(d \theta) 
\label{bdgen}
\eeq
Therefore, we have the recursion relation
\beq 
b^{(d+2)}+(-1)^{d+1}b^{(1)} = (t+\frac{1}{t})(b^{(d+1)}+(-1)^d b^{(1)}) -
(b^{(d)}+(-1)^{d-1}b^{(1)}) \qquad \mbox{for} \ d \ge 2 \ , \eeq
that is
\beq
b^{(d+2)} = (q-2)b^{(d+1)} - b^{(d)} + q(-1)^d b^{(1)} \qquad \mbox{for} \ 
d \ge 2 \ .
\eeq
Now the coefficient $c^{(d)}$ in \cite{cf} is given by
\beq
c^{(d)} = U_{2d}(\frac{\sqrt{q}}{2}) = \sum_{j=0}^d (-1)^j {2d-j \choose j}
q^{d-j}
\label{cd}
\eeq
where $U_n(x)$ is the Chebyshev polynomial of the second kind.  Equivalently,
in terms of the angle $\theta$ in eq. (\ref{qtheta}) \cite{saleur,cf}, 
\beq
c^{(d)} = \frac{\sin \left ( (2d+1)\frac{\theta}{2} \right )}
               {\sin \left ( \frac{\theta}{2} \right )}
\label{cdequiv}
\eeq
In general, we find that the relation between $b^{(d)}$ and $c^{(d)}$ is as 
follows,
\beqs
c^{(d)} & = & b^{(d)} \qquad \mbox{for} \ 0 \le d \le 2 \cr\cr
c^{(d)} & = & \sum _{d^\prime=1}^d b^{(d^\prime)} \qquad \mbox{for odd} \ d 
\cr\cr
c^{(d)} & = & \sum _{d^\prime=2}^d b^{(d^\prime)} \qquad \mbox{for even} \ d 
\ ,
\eeqs
and
\beqs
b^{(d)} & = & c^{(d)} - c^{(d-1)} + (-1)^d c^{(1)} \cr\cr
& = & U_{2d}(\frac{\sqrt{q}}{2}) - U_{2d-2}(\frac{\sqrt{q}}{2}) + (-1)^d (q-1)
 \cr\cr 
& = & q^d + \sum_{j=1}^d (-1)^j \frac{2d}{2d-j} {2d-j \choose j} q^{d-j} + 
(-1)^d (q-1) \qquad \mbox{for} \ d \ge 2
\eeqs
At the value $q=0$, $b^{(d)}$ has the property that
\beq
b^{(d)}(q=0) = (-1)^d  \ .
\label{bdq0}
\eeq
We also find the following properties of the $b^{(d)}$ polynomials. 
For $q=1$ and $d \ge 2$, $b^{(d)}$ is equal to 2 if $d=0$ mod 3 and
$-1$ if $d=1$ or $d=2$ mod 3.  For $q=2$ and $d \ge 2$, $b^{(d)}$ is equal
to 3 if $d=0$ mod 4 and $-1$ if $d=1,2,3$ mod 4.  One can derive similar
relations for other values of $q$.  For $q=3$, $b^{(d)}$ takes on the
pattern of values $-4,1,-1,4,-1,1$ for $d=3,4,5,6,7,8$, and then this pattern
repeats for higher integral values of $d$.  For $q=4$ and $d \ge 2$,
$b^{(d)}=5$ if $d$ is even and $b^{(d)}=-1$ if $d$ is odd. For larger values of
$q$, $b^{(d)}$ does not cycle through a fixed pattern of values as $d$
increases, but instead increases.  The fixed patterns of values of $b^{(d)}$
for integer $q$ from 0 to 4 are related to special sum rules for Potts model
partition functions on lattice strips with toroidal boundary conditions,
similar to those that we exhibited in Refs. \cite{ka3} and \cite{s3a}.

\begin{figure}[htbp]
\unitlength 1mm \hspace*{5mm}
\begin{picture}(115,4)
\put(10,4){\makebox(0,0){{\small $d=2$}}}
\multiput(20,0)(0,4){2}{\circle{2}} \put(20,0){\line(0,1){4}}
\put(25,2){\makebox(0,0){{\small $-$}}}
\put(30,2){\makebox(0,0){{\small $2$}}} \put(35,2){\circle{2}}
\put(40,2){\makebox(0,0){{\small $+$}}}
\put(45,2){\makebox(0,0){{\small $1$}}}
\put(65,4){\makebox(0,0){{\small $d=3$}}}
\multiput(75,0)(4,0){2}{\circle{2}} \put(75,0){\line(1,0){4}}
\put(77,4){\circle{2}} \put(75,0){\line(1,2){2}} \put(79,0){\line(-1,2){2}}
\put(85,2){\makebox(0,0){{\small $-$}}}
\put(90,2){\makebox(0,0){{\small $3$}}}
\multiput(95,0)(0,4){2}{\circle{2}} \put(95,0){\line(0,1){4}}
\put(100,2){\makebox(0,0){{\small $+$}}}
\put(105,2){\makebox(0,0){{\small $3$}}} \put(110,2){\circle{2}}
\put(115,2){\makebox(0,0){{\small $-$}}}
\put(120,2){\makebox(0,0){{\small $1$}}}
\end{picture}

\vspace*{10mm} \hspace*{-5mm}
\begin{picture}(105,8)
\put(10,8){\makebox(0,0){{\small $d=4$}}}
\multiput(20,2)(0,4){2}{\circle{2}} \put(20,2){\line(0,1){4}}
\multiput(24,2)(0,4){2}{\circle{2}} \put(24,2){\line(0,1){4}}
\put(20,2){\line(1,0){4}} \put(20,6){\line(1,0){4}}
\put(30,4){\makebox(0,0){{\small $-$}}}
\put(35,4){\makebox(0,0){{\small $4$}}}
\multiput(40,0)(0,4){3}{\circle{2}} \put(40,0){\line(0,1){8}}
\put(44,4){\makebox(0,0){{\small $+$}}}
\put(50,4){\makebox(0,0){{\small $($}}}
\put(54,4){\makebox(0,0){{\small $4$}}}
\multiput(60,2)(0,4){2}{\circle{2}} \put(60,2){\line(0,1){4}}
\put(65,4){\makebox(0,0){{\small $+$}}}
\put(70,4){\makebox(0,0){{\small $2$}}}
\multiput(75,2)(0,4){2}{\circle{2}} 
\put(80,4){\makebox(0,0){{\small $)$}}}
\put(85,4){\makebox(0,0){{\small $-$}}}
\put(90,4){\makebox(0,0){{\small $4$}}}
\put(95,4){\circle{2}}
\put(100,4){\makebox(0,0){{\small $+$}}}
\put(105,4){\makebox(0,0){{\small $1$}}}
\end{picture}

\caption{\footnotesize{Multiplicities $b^{(d)}$ for $2 \le d \le 4$.}} \label{bdfigure}
\end{figure}

The partitions for toroidal strips are more involved than those for cyclic
strips because of the rotational symmetry of the slice of toroidal strips. We
list graphically all the possible partitions for strips of widths $L_y=2$ and
$L_y=3$ in Figs. \ref{L2partitions} and \ref{L3partitions}, respectively, where
white circles are the original $L_y$ vertices and each black circle corresponds
to a specific color assignment. In the following discussion, we will simply use
the names white and black circles with these meanings understood
implicitly. The connections between the black circles and white circles should
obey the non-crossing restriction. The apparent crossings in some partitions in
Figs. \ref{L2partitions} and \ref{L3partitions} do not violate this rule when
white circles are rotated appropriately, owing to the rotational symmetry 
for toroidal strips. For example, the crossing in the last
partition of $d=1$ in Fig. \ref{L3partitions} can be removed by moving the
bottom white circle to the top so that it becomes equivalent to the fourth
partition of $d=1$, or moving the top white circle to the bottom so that it
becomes equivalent to the eighth partition of $d=1$ in the figure. Another
example is the crossing in the eleventh partition of $d=2$ in
Fig. \ref{L3partitions} that can be removed by the rotational symmetry of
the white circles.  In addition, it is necessary to include all the
arrangements of black circles when the transfer matrix is constructed. For
example, the two partitions of $d=2$ in Fig. \ref{L2partitions} correspond to
exchanging the black circles, i.e. the assignments of the colors.  We arrange
the partitions of $d=2$ in Fig. \ref{L3partitions} into pairs such that every
two partitions are equivalent when the two black circles are exchanged.  We
show the six partitions of $d=3$ in Fig. \ref{L3partitions} which correspond to
all the possible arrangements of black circles.

\begin{figure}
\unitlength 1mm \hspace*{5mm}
\begin{picture}(120,12)
\put(10,12){\makebox(0,0){{\small $d=0$}}}
\multiput(20,8)(10,0){2}{\circle{2}}
\multiput(20,12)(10,0){2}{\circle{2}} \put(30,8){\line(0,1){4}}
\put(50,12){\makebox(0,0){{\small $d=1$}}}
\multiput(60,4)(10,0){3}{\circle*{2}}
\multiput(60,8)(10,0){3}{\circle{2}}
\multiput(60,12)(10,0){3}{\circle{2}} \put(60,4){\line(0,1){4}}
\put(70,8){\oval(4,8)[l]} \put(80,4){\line(0,1){8}}
\put(100,12){\makebox(0,0){{\small $d=2$}}}
\multiput(110,0)(0,4){2}{\circle*{2}}
\multiput(110,8)(0,4){2}{\circle{2}} \put(110,4){\line(0,1){4}}
\put(110,6){\oval(4,12)[l]}
\multiput(120,0)(0,4){2}{\circle*{2}}
\multiput(120,8)(0,4){2}{\circle{2}} \put(120,4){\oval(4,8)[l]}
\put(120,8){\oval(6,8)[l]}
\end{picture}

\caption{\footnotesize{Partitions for the $L_y=2$ strip.}}
\label{L2partitions}
\end{figure}
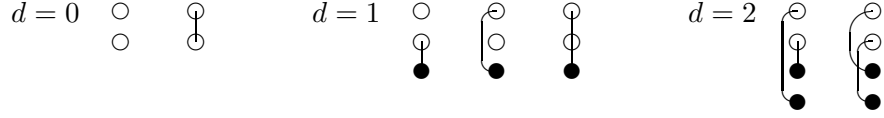

\begin{figure}
\unitlength 1mm \hspace*{5mm}
\begin{picture}(60,8)
\put(10,8){\makebox(0,0){{\small $d=0$}}}
\multiput(20,0)(10,0){5}{\circle{2}}
\multiput(20,4)(10,0){5}{\circle{2}}
\multiput(20,8)(10,0){5}{\circle{2}} \put(30,4){\line(0,1){4}}
\put(40,4){\oval(4,8)[l]} \put(50,0){\line(0,1){4}}
\put(60,0){\line(0,1){8}}
\end{picture}

\vspace*{5mm} \hspace*{5mm}
\begin{picture}(110,12)
\put(10,12){\makebox(0,0){{\small $d=1$}}}
\multiput(20,0)(10,0){10}{\circle*{2}}
\multiput(20,4)(10,0){10}{\circle{2}}
\multiput(20,8)(10,0){10}{\circle{2}}
\multiput(20,12)(10,0){10}{\circle{2}} \put(20,0){\line(0,1){4}}
\put(30,4){\oval(4,8)[l]} \put(40,6){\oval(4,12)[l]}
\put(50,0){\line(0,1){4}} \put(50,8){\line(0,1){4}}
\put(60,4){\oval(4,8)[l]} \put(60,8){\line(0,1){4}}
\put(70,0){\line(0,1){4}} \put(70,8){\oval(4,8)[l]}
\put(80,0){\line(0,1){8}} \put(90,6){\oval(4,12)[l]}
\put(90,4){\line(0,1){4}} \put(100,0){\line(0,1){12}}
\put(110,4){\oval(4,8)[l]} \put(110,8){\oval(6,8)[l]}
\end{picture}

\vspace*{5mm} \hspace*{5mm}
\begin{picture}(130,16)
\put(10,16){\makebox(0,0){{\small $d=2$}}}
\multiput(20,0)(10,0){12}{\circle*{2}}
\multiput(20,4)(10,0){12}{\circle*{2}}
\multiput(20,8)(10,0){12}{\circle{2}}
\multiput(20,12)(10,0){12}{\circle{2}}
\multiput(20,16)(10,0){12}{\circle{2}} 
\put(20,4){\line(0,1){4}} \put(20,6){\oval(4,12)[l]} 
\put(30,4){\oval(4,8)[l]} \put(30,8){\oval(6,8)[l]}
\put(40,4){\line(0,1){4}} \put(40,8){\oval(4,16)[l]} 
\put(50,4){\oval(4,8)[l]} \put(50,10){\oval(6,12)[l]}
\put(60,8){\oval(4,8)[l]} \put(60,8){\oval(6,16)[l]} 
\put(70,6){\oval(4,12)[l]} \put(70,10){\oval(6,12)[l]}
\put(80,4){\line(0,1){4}} \put(80,6){\oval(4,12)[l]} \put(80,12){\line(0,1){4}}
\put(90,4){\oval(4,8)[l]} \put(90,8){\oval(6,8)[l]} \put(90,12){\line(0,1){4}}
\put(100,4){\line(0,1){8}} \put(100,8){\oval(4,16)[l]}
\put(110,4){\oval(4,8)[l]} \put(110,10){\oval(6,12)[l]} \put(110,8){\line(0,1){4}}
\put(120,6){\oval(4,12)[l]} \put(120,12){\oval(6,8)[l]} \put(120,4){\line(0,1){4}}
\put(130,4){\oval(4,8)[l]} \put(130,12){\oval(4,8)[l]} \put(130,8){\oval(6,8)[l]}
\end{picture}

\vspace*{5mm} \hspace*{5mm}
\begin{picture}(70,20)
\put(10,20){\makebox(0,0){{\small $d=3$}}}
\multiput(20,0)(10,0){6}{\circle*{2}}
\multiput(20,4)(10,0){6}{\circle*{2}}
\multiput(20,8)(10,0){6}{\circle*{2}}
\multiput(20,12)(10,0){6}{\circle{2}}
\multiput(20,16)(10,0){6}{\circle{2}} 
\multiput(20,20)(10,0){6}{\circle{2}} 
\put(20,8){\line(0,1){4}} \put(20,10){\oval(4,12)[l]} \put(20,10){\oval(6,20)[l]}
\put(30,8){\line(0,1){4}} \put(30,8){\oval(4,16)[l]} \put(30,12){\oval(6,16)[l]}
\put(40,12){\oval(4,8)[l]} \put(40,8){\oval(6,8)[l]} \put(40,10){\oval(8,20)[l]}
\put(50,12){\oval(4,8)[l]} \put(50,12){\oval(6,16)[l]} \put(50,6){\oval(8,12)[l]}
\put(60,14){\oval(4,12)[l]} \put(60,8){\oval(6,8)[l]} \put(60,8){\oval(8,16)[l]}
\put(70,14){\oval(4,12)[l]} \put(70,10){\oval(6,12)[l]} \put(70,6){\oval(8,12)[l]}
\end{picture}

\caption{\footnotesize{Partitions for the $L_y=3$ strip.}}
\label{L3partitions}
\end{figure}

We denote the partitions ${\cal P}_{L_y,d}$ for $2 \le L_y \le 4$ as follows:
\beq 
{\cal P}_{2,0} = \{ I; 12 \} \ , \qquad 
{\cal P}_{2,1} = \{ \bar 2; \bar 1; \overline{12} \} \ , \qquad 
{\cal P}_{2,2} = \{ \bar 1, \hat 2 ; \hat 1, \bar 2 \} 
\label{L2partitionlist} 
\eeq
\beqs 
{\cal P}_{3,0} & = & \{ I; 12; 13; 23; 123 \} \ , \qquad 
{\cal P}_{3,1} = \{ \bar 3; \bar 2; \bar 1; 12, \bar 3; \overline{12}; \overline{13}; \overline{23}; 23, \bar 1; \overline{123}; 13, \bar 2 \} \ , \cr\cr 
{\cal P}_{3,2} & = & \{ \bar 2, \hat 3; \hat 2, \bar 3; \bar 1, \hat 3; \hat 1, \bar 3; \bar 1, \hat 2; \hat 1, \bar 2; \overline{12}, \hat 3; \widehat{12}, \bar 3; \bar 1, \widehat{23}; \hat 1, \overline{23}; \overline{13}, \hat 2; \widehat{13}, \bar 2 \} \ , \cr\cr
{\cal P}_{3,3} & = & \{ \bar 1, \hat 2, \tilde 3; \hat 1, \bar 2, \tilde 3; \bar 1, \tilde 2, \hat 3; \hat 1, \tilde 2, \bar 3; \tilde 1, \bar 2, \hat 3; \tilde 1, \hat 2, \bar 3 \} 
\label{L3partitionlist} 
\eeqs
\beqs 
{\cal P}_{4,0} & = & \{ I; 12; 13; 14; 23; 24; 34; 12, 34; 14, 23; 123;
124; 134; 234; 1234 \} \ , \cr\cr 
{\cal P}_{4,1} & = & \{ \bar 4; \bar 3; \bar 2; \bar 1; 12, \bar 4; 12, \bar 3; \overline{12}; 13, \bar 4; \overline{13}; \overline{14}; 23, \bar 4; \overline{23}; 23, \bar 1; \overline{24}; 24, \bar 1; \overline{34}; 34, \bar 2; 34, \bar 1; 12, \overline{34}; \cr\cr 
& & 34, \overline{12}; 23, \overline{14}; 123, \bar 4; \overline{123}; \overline{124};
\overline{134}; \overline{234}; 234, \bar 1; \overline{1234}; 13, \bar 2; 14, \bar 3; 14, \bar 2; 24, \bar 3; 14, \overline{23}; \cr\cr
& & 124, \bar 3; 134, \bar 2 \} \ , \cr\cr
{\cal P}_{4,2} & = & \{ \bar 3, \hat 4; \hat 3, \bar 4; \bar 2, \hat 4; \hat 2, \bar 4; \bar 1, \hat 4; \hat 1, \bar 4; \bar 2, \hat 3; \hat 2, \bar 3; \bar 1, \hat 3; \hat 1, \bar 3; \bar 1, \hat 2; \hat 1, \bar 2; 12, \bar 3, \hat 4; 12, \hat 3, \bar 4; \overline{12}, \hat 4; \widehat{12}, \bar 4; \cr\cr
& & \overline{12}, \hat 3; \widehat{12}, \bar 3; \overline{13}, \hat 4; \widehat{13}, \bar 4; \overline{13}, \hat 2; \widehat{13}, \bar 2; \overline{14}, \hat 3; \widehat{14}, \bar 3; \overline{14}, \hat 2; \widehat{14}, \bar 2; 14, \bar 2, \hat 3; 14, \hat 2, \bar 3; \overline{23}, \hat 4; \widehat{23}, \bar 4; \cr\cr
& & 23, \bar 1, \hat 4; 23, \hat 1, \bar 4; \bar 1, \widehat{23}; \hat 1, \overline{23}; \bar 1, \widehat{24}; \hat 1, \overline{24}; \overline{24}, \hat 3; \widehat{24}, \bar 3; \bar 2, \widehat{34}; \hat 2, \overline{34}; \bar 1, \widehat{34}; \hat 1, \overline{34}; 34, \bar 1, \hat 2; \cr\cr
& & 34, \hat 1, \bar 2; \overline{12}, \widehat{34}; \hat{12}, \overline{34}; \overline{14}, \widehat{23}; \widehat{14}, \overline{23}; \overline{123}, \hat 4; \widehat{123}, \bar 4; \overline{124}, \hat 3; \widehat{124}, \bar 3; \overline{134}, \hat 2; \widehat{134}, \bar 2; \bar 1, \widehat{234}; \cr\cr
& & \hat 1, \overline{234} \} \ , \cr\cr 
{\cal P}_{4,3} & = & \{ \bar 2, \hat 3, \tilde 4; \hat 2, \bar 3, \tilde 4; \bar 2, \tilde 3, \hat 4; \hat 2, \tilde 3, \bar 4; \tilde 2, \bar 3, \hat 4; \tilde 2, \hat 3, \bar 4; 
\bar 1, \hat 3, \tilde 4; \hat 1, \bar 3, \tilde 4; \bar 1, \tilde 3, \hat 4; \hat 1, \tilde 3, \bar 4; \tilde 1, \bar 3, \hat 4; \tilde 1, \hat 3, \bar 4; \cr\cr
& & \bar 1, \hat 2, \tilde 4; \hat 1, \bar 2, \tilde 4; \bar 1, \tilde 2, \hat 4; \hat 1, \tilde 2, \bar 4; \tilde 1, \bar 2, \hat 4; \tilde 1, \hat 2, \bar 4;
\bar 1, \hat 2, \tilde 3; \hat 1, \bar 2, \tilde 3; \bar 1, \tilde 2, \hat 3; \hat 1, \tilde 2, \bar 3; \tilde 1, \bar 2, \hat 3; \tilde 1, \hat 2, \bar 3; \cr\cr
& & \overline{12}, \hat 3, \tilde 4; \widehat{12}, \bar 3, \tilde 4; \overline{12}, \tilde 3, \hat 4; \widehat{12}, \tilde 3, \bar 4; \widetilde{12}, \bar 3, \hat 4; \widetilde{12}, \hat 3, \bar 4; 
\overline{14}, \hat 2, \tilde 3; \widehat{14}, \bar 2, \tilde 3; \overline{14}, \tilde 2, \hat 3; \widehat{14}, \tilde 2, \bar 3; \cr\cr & & \widetilde{14}, \bar 2, \hat 3; \widetilde{14}, \hat 2, \bar 3;
\bar 1, \widehat{23}, \tilde 4; \hat 1, \overline{23}, \tilde 4; \bar 1, \widetilde{23}, \hat 4; \hat 1, \widetilde{23}, \bar 4; \tilde 1, \overline{23}, \hat 4; \tilde 1, \widehat{23}, \bar 4;  
\bar 1, \hat 2, \widetilde{34}; \hat 1, \bar 2, \widetilde{34}; \cr\cr & & \bar 1, \tilde 2, \widehat{34}; \hat 1, \tilde 2, \overline{34}; \tilde 1, \bar 2, \widehat{34}; \tilde 1, \hat 2, \overline{34} \} \ , \cr\cr
{\cal P}_{4,4} & = & \{ \bar 1, \hat 2, \check 3, \tilde 4; \hat 1, \bar 2, \check 3, \tilde 4; \bar 1, \check 2, \hat 3, \tilde 4; \hat 1, \check 2, \bar 3, \tilde 4; \check 1, \bar 2, \hat 3, \tilde 4; \check 1, \hat 2, \bar 3, \tilde 4;
\bar 1, \hat 2, \tilde 3, \check 4; \hat 1, \bar 2, \tilde 3, \check 4; \bar 1, \check 2, \tilde 3, \hat 4; \cr\cr & & \hat 1, \check 2, \tilde 3, \bar 4; \check 1, \bar 2, \tilde 3, \hat 4; \check 1, \hat 2, \tilde 3, \bar 4;
\bar 1, \tilde 2, \hat 3, \check 4; \hat 1, \tilde 2, \bar 3, \check 4; \bar 1, \tilde 2, \check 3, \hat 4; \hat 1, \tilde 2, \check 3, \bar 4; \check 1, \tilde 2, \bar 3, \hat 4; \check 1, \tilde 2, \hat 3, \bar 4; \cr\cr & &
\tilde 1, \bar 2, \hat 3, \check 4; \tilde 1, \hat 2, \bar 3, \check 4; \tilde 1, \bar 2, \check 3, \hat 4; \tilde 1, \hat 2, \check 3, \bar 4; \tilde 1, \check 2, \bar 3, \hat 4; \tilde 1, \check 2, \hat 3, \bar 4 \}
\label{L4partitionlist} 
\eeqs
where partitions are separated by a colon, and overline, hat, check and tilde
in each partition correspond to different color assignments. It follows that
the size of the transfer matrix $T_{Z,\Lambda,L_y,d}$ always has the factor
$d!$. Let us denote the reduced size of the transfer matrix as
$n_{Z,tor}(L_y,d)$, i.e., we neglect the permutation of black circles.  As the
derivation of $b^{(d)}$ does not take into account permutations, i.e. a set of
$d!$ eigenvalues should have their coefficients summed to be equal to
$b^{(d)}$, we find that the sum of all coefficients for a toroidal strip graph,
which is equal to the dimension of the total transfer matrix, i.e., 
the number of ways to color $L_y$ vertices, is 
\beq 
C_{Z,\Lambda,L_y} = dim(T_{Z,\Lambda,L_y}) = \sum _{d=0}^{L_y}
n_{Z,tor}(L_y,d) b^{(d)} = q^{L_y} \qquad {\rm for} \ \ \Lambda=sq,tri,hc \ .
\label{czsumtor}
\eeq

By substituting the expression for $b^{(d)}$ in eq. (\ref{czsumtor}), we
determine the $n_{Z,tor}(L_y,d)$, as follows: 
\beq
n_{Z,tor}(L_y,d) = 0 \qquad \mbox{for} \ d>L_y
\label{nzup}
\eeq
\beq
n_{Z,tor}(L_y,0) = C_{L_y}
\label{nzly0}
\eeq
\beq
n_{Z,tor}(L_y,1) = {2L_y-1 \choose L_y-1} 
\label{nzly1}
\eeq
\beq
n_{Z,tor}(L_y,d) = {2L_y \choose L_y-d} \qquad \mbox{for} \ 2 \le d \le L_y
\label{nzlyd}
\eeq
where $C_n$ is the Catalan number which occurs in combinatorics and is 
defined by
\beq
C_n = \frac{1}{(n+1)}{2n \choose n} \ . 
\label{catalan}
\eeq
The first few Catalan numbers are $C_1=1$, $C_2=2$, $C_3=5$, $C_4=14$, and
$C_5=42$. Special cases of $n_{Z,tor}(L_y,d)$ that are of interest here include
\beq
n_{Z,tor}(L_y,L_y)=1
\label{nzlyly}
\eeq
\beq
n_{Z,tor}(L_y,L_y-1)=2L_y \qquad \mbox{for} \ L_y \ge 3
\label{nzlylyminus1}
\eeq
In Table \ref{nztabletor} we list the first few numbers $n_{Z,tor}(L_y,d)$. We
notice the following recursion relations:
\beqs
n_{Z,tor}(L_y,1) & = & n_{Z,tor}(L_y,0) - n_{Z,tor}(L_y-1,0) + 2n_{Z,tor}(L_y-1,1) + n_{Z,tor}(L_y-1,2) \cr\cr
& & \qquad \mbox{for} \ L_y \ge 2 \cr\cr
n_{Z,tor}(L_y,2) & = & -n_{Z,tor}(L_y-1,0) + 2n_{Z,tor}(L_y-1,1) + 2n_{Z,tor}(L_y-1,2) + n_{Z,tor}(L_y-1,3) \cr\cr
& & \qquad \mbox{for} \ L_y \ge 2 \cr\cr
n_{Z,tor}(L_y,d) & = & n_{Z,tor}(L_y-1,d-1) + 2n_{Z,tor}(L_y-1,d) + n_{Z,tor}(L_y-1,d+1) \cr\cr
& & \qquad \mbox{for} \ L_y \ge 3, \ 3 \le d \le L_y \ . 
\eeqs
The last line here is the same as the recursion relation for $n_{Z,cyc}(L_y,d)$
when $1 \le d \le L_y$ for cyclic strips (which was denoted as $n_Z(L_y,d)$ in
\cite{cf}). By substituting $q=0$ into eq. (\ref{czsumtor}) and using
eq. (\ref{bdq0}), we have the relation
\beq
\sum _{d=0}^{L_y} n_{Z,tor}(L_y,d) (-1)^{d} = 0 \qquad {\rm for} \ \ \Lambda=sq,tri,hc \ . 
\eeq

\begin{table}[htbp]
\caption{\footnotesize{Table of numbers $n_{Z,tor}(L_y,d)$.  See text for general formulas.}}
\begin{center}
\begin{tabular}{|c|c|c|c|c|c|c|c|c|c|c|c|}
\hline\hline $L_y \backslash \ d$ 
   & 0   & 1   & 2    & 3   & 4   & 5   & 6  & 7  & 8 & 9 & 10 \\ \hline\hline
1  & 1   & 1   &      &     &     &     &    &    &   &   &    \\ \hline
2  & 2   & 3   & 1    &     &     &     &    &    &   &   &    \\ \hline
3  & 5   & 10  & 6    & 1   &     &     &    &    &   &   &    \\ \hline
4  & 14  & 35  & 28   & 8   & 1   &     &    &    &   &   &    \\ \hline
5  & 42  & 126 & 120  & 45  & 10  &  1  &    &    &   &   &    \\ \hline
6  & 132 & 462 & 495  & 220 & 66  &  12 & 1  &    &   &   &    \\ \hline
7  & 429 & 1716& 2002 & 1001& 364 &  91 & 14 &  1 &   &   &    \\ \hline
8  & 1430& 6435& 8008 & 4368& 1820&  560& 120&  16& 1 &   &    \\ \hline
9  & 4862&24310& 31824&18564& 8568& 3060& 816& 153& 18& 1 &    \\ \hline
10 &16796&92378&125970&77520&38760&15504&4845&1140&190&20 & 1  \\ \hline\hline
\end{tabular}
\end{center}
\label{nztabletor}
\end{table}

Another way to realize the number $n_{Z,tor}(L_y,d)$ is as follows. For $d=0$,
i.e. no color assignment, the partitions are identical to those for cyclic
strips, and the number of these partitions is 
$n_{Z,tor}(L_y,0)=n_{Z,cyc}(L_y,0)=C_{L_y}$ given by eq. (\ref{nzly0}).  Recall
that the number of partitions of size $n$ with $k$ components is given by the
Narayana number $N(n,k)$, which is sequence A001263 in \cite{sl} (see also
Ref. \cite{Flajolet99} and the references therein), given by
\beq 
N(n,k) = \frac{1}{k} {n-1 \choose k-1} {n \choose k-1} = \frac{1}{n} {n
\choose k} {n \choose k-1} \ .  \eeq
The first few rows of the triangle of Narayana numbers, or Catalan triangle,
are shown in Fig. \ref{narayanatriangle}. It is clear that the sum of terms in 
each row in the triangle is equal to the corresponding Catalan number:
\beq
\sum _{k=1}^{L_y} N(L_y,k) = \frac{1}{L_y} \sum_{k=0}^{L_y-1} {L_y \choose k+1} {L_y \choose k} = \frac{1}{L_y} {2L_y \choose L_y-1} = C_{L_y} =n_{Z,tor}(L_y,0) \ .
\eeq
For $d=1$, we need to assign a color for each component of the level 0
partitions.  Therefore, we have
\beqs 
n_{Z,tor}(L_y,1) & = & \sum_{k=1}^{L_y} kN(L_y,k) = \sum_{k=1}^{L_y}
{L_y-1 \choose k-1} {L_y \choose k-1} = {2L_y-1 \choose L_y-1} \cr\cr & = &
\frac{1}{2} {2L_y \choose L_y} = \frac{L_y+1}{2}C_{L_y} =
\frac{L_y+1}{2}n_{Z,tor}(L_y,0) \ .  \eeqs
Let us compare $n_{Z,tor}(L_y,d)$ with the corresponding number
$n_{Z,cyc}(L_y,d)$ given in Table 3 of \cite{cf}. In general,
$n_{Z,tor}(L_y,d)$ is larger than $n_{Z,cyc}(L_y,d)$, except for $d=0$ where
they are the same. This is a consequence of the fact that certain forbidden
partitions for cyclic strips for $d \ge 1$ are allowed for toroidal strips with
rotational translation. We list the first few
$n_{Z,tor}(L_y,d)-n_{Z,cyc}(L_y,d)$ in Table \ref{nztabletorcyc}. Comparing
Table \ref{nztabletorcyc} with Table \ref{nztabletor}, we infer the relation
\beq n_{Z,tor}(L_y,d) - n_{Z,cyc}(L_y,d) = n_{Z,tor}(L_y,d+1) \qquad \mbox{for}
\ d \ge 2 \ .
\label{nztorcyc}
\eeq
This can be understood as follows. The extra partitions for toroidal strips, as
contrast with cyclic strips, are those with removable crossings. That is, for
any connection between two non-adjacent white circles, all the color
assignments should be inside or outside the connection while these two white
circles can be either color-assigned or not. Therefore, there are three
possible correspondences from the partitions for $n_{Z,tor}(L_y,d+1)$ to the
partitions for $n_{Z,tor}(L_y,d) - n_{Z,cyc}(L_y,d)$ with $d \ge 2$: (i) if the
top color-assigned vertex and the bottom color-assigned vertex are not within a
connection, then connect these two vertices (so that only one color is assigned
to them); (ii) if all color-assigned vertices are within a connection and the
connection is also assigned a color, then remove the assignment for the
connection; (iii) if all color-assigned vertices are within a connection and
the connection is not assigned a color, then connect the top and bottom
color-assigned vertices (within that connection). These account for all of the
extra partitions that toroidal strips have, relative to cyclic strips. As an
example, consider the case of width $L_y=4$ and degree $d=2$ in
eq. (\ref{L4partitionlist}). For this case there are eight partitions that are
allowed for toroidal strips but not for cyclic strips.  These are 
$\overline{13}, \bar 2$; $\overline{14}, \bar 3$; $\overline{14}, \bar 2$; $14,
\bar 2, \bar 3$; $\overline{24}, \bar 3$; $\overline{14}, \overline{23}$;
$\overline{124}, \bar 3$; $\overline{134}, \bar 2$. Here we only use overline
to indicate the color assignment, since permutations of black circles are not
considered for $n_{Z,tor}(L_y,d)$. These partitions are in one-to-one
correspondence with the partitions for $L_y=4$ and $d=3$ (without color
permutation), namely $\bar 1, \bar 2, \bar 3$; $\bar 1, \bar 3, \bar 4$; $\bar
1, \bar 2, \bar 4$; $\overline{14}, \bar 2, \bar 3$; $\bar 2, \bar 3, \bar 4$;
$\overline{23}, \bar 1, \bar 4$; $\overline{12}, \bar 3, \bar 4$;
$\overline{34}, \bar 1, \bar 2$. Eq. (\ref{nztorcyc}) is equivalent to the
relation
\beq
n_{Z,tor}(L_y,d) = \sum_{d^\prime=d}^{L_y} n_{Z,cyc}(L_y,d^\prime) \qquad \mbox{for} \ d \ge 2 \ .
\eeq

\begin{figure}[htbp]
\begin{center}
\begin{tabular}{ccccccccccccccccc}
  &   &   &   &    &    &    &     & 1   &     &     &    &    &   &   &   &   \\
  &   &   &   &    &    &    & 1   &     & 1   &     &    &    &   &   &   &   \\
  &   &   &   &    &    & 1  &     & 3   &     & 1   &    &    &   &   &   &   \\
  &   &   &   &    & 1  &    & 6   &     & 6   &     & 1  &    &   &   &   &   \\
  &   &   &   & 1  &    & 10 &     & 20  &     & 10  &    & 1  &   &   &   &   \\
  &   &   & 1 &    & 15 &    & 50  &     & 50  &     & 15 &    & 1 &   &   &   \\
  &   & 1 &   & 21 &    & 105&     & 175 &     & 105 &    & 21 &   & 1 &   &   \\
  & 1 &   & 28&    & 196&    & 490 &     & 490 &     & 196&    & 28&   & 1 &   \\
1 &   & 36&   & 336&    &1176&     &1764&      &1176 &    & 336&   & 36&   & 1
\end{tabular}
\end{center}
\caption{\footnotesize{Triangle of Narayana numbers.}}
\label{narayanatriangle}
\end{figure}

\begin{table}[htbp]
\caption{\footnotesize{Table of numbers $n_{Z,tor}(L_y,d)-n_{Z,cyc}(L_y,d)$.}}
\begin{center}
\begin{tabular}{|c|c|c|c|c|c|c|c|c|c|c|c|}
\hline\hline $L_y \backslash \ d$ 
   & 0 & 1   & 2   & 3   & 4   & 5  & 6  & 7 & 8 & 9 & 10 \\ \hline\hline
1  & 0 & 0   &     &     &     &     &   &   &   &   &    \\ \hline
2  & 0 & 0   & 0   &     &     &     &   &   &   &   &    \\ \hline
3  & 0 & 1   & 1   & 0   &     &     &   &   &   &   &    \\ \hline
4  & 0 & 7   & 8   & 1   & 0   &     &   &   &   &   &    \\ \hline
5  & 0 & 36  & 45  & 10  & 1   & 0   &   &   &   &   &    \\ \hline
6  & 0 & 165 & 220 & 66  & 12  & 1  & 0  &   &   &   &    \\ \hline
7  & 0 & 715 & 1001& 364 & 91  & 14 & 1  & 0 &   &   &    \\ \hline
8  & 0 & 3003& 4368& 1820& 560 & 120& 16 & 1 & 0 &   &    \\ \hline
9  & 0 &12376&18564& 8568& 3060& 816& 153& 18& 1 & 0 &    \\ \hline
10 & 0 &50388&77520&38760&15504&4845&1140&190& 20& 1 & 0  \\ \hline\hline  
\end{tabular}
\end{center}
\label{nztabletorcyc}
\end{table}

The construction of the transfer matrix for each level (= degree) $d$ can be
carried out by methods similar to those for cyclic strips \cite{zt}. Notice
that for the honeycomb lattice, the number of vertices in the transverse
direction, $L_y$, should be an even number, and the smallest value without
degeneracy is $L_y=4$. Using the bases described above
(e.g. eqs. (\ref{L2partitionlist})-(\ref{L4partitionlist}) for $2 \le L_y \le
4$), we define $J_{L_y,d,i,i+1}$ as the join operator between vertices $i$ and
$i+1$ and $D_{L_y,d,i}$ as the detach operator on vertex $i$ for each subspace
$d$.  As compared with the situation for cyclic strips, the corresponding
toroidal strips with the same $L_y$ and $d$ have one more join operator, namely
$J_{L_y,d,L_y,1} \equiv J_{L_y,d,L_y,L_y+1}$.  The transfer matrix
$T_{Z,\Lambda,L_y,d}$ for each $d$ is the product of the transverse and
longitudinal parts, $H_{Z,\Lambda,L_y,d}$ and $V_{Z,\Lambda,L_y,d}$, which can
be expressed as
\beqs 
H_{Z,sq,L_y,d} & = & \prod_{i=1}^{L_y} (I+vJ_{L_y,d,i,i+1}) \cr\cr 
H_{Z,tri,L_y,d} & = & J_{L_y+1,d,L_y+1,1} \prod_{i=1}^{L_y} (I+vJ_{L_y+1,d,i,i+1}) \cr\cr  
H_{Z,hc,L_y,d,1} & = & \prod _{i=1}^{L_y/2} (I+vJ_{L_y,d,2i-1,2i}) \ , \qquad 
H_{Z,hc,L_y,d,2} = \prod _{i=1}^{L_y/2} (I+vJ_{L_y,d,2i,2i+1}) \cr\cr 
V_{Z,sq,L_y,d} & = & V_{Z,hc,L_y,d} = \prod_{i=1}^{L_y} (vI+D_{L_y,d,i}) \cr\cr
V_{Z,tri,L_y,d} & = & D_{L_y+1,d,1} \prod_{i=1}^{L_y} [(I+vJ_{L_y+1,d,i,i+1}) (vI+D_{L_y+1,d,i+1})] \ , 
\label{HVmatrix} 
\eeqs
where $[\nu]$ denotes the integral part of $\nu$, and $I$ is the identity
matrix. Notice that for triangular strips with periodic transverse boundary
conditions, one has to work with width-$(L_y+1)$ matrices and then identify the
two end vertices, as shown in \cite{sqtran}. We have
\beqs 
T_{Z,sq,L_y,d} & = & V_{Z,sq,L_y,d} H_{Z,sq,L_y,d} \ , \qquad
T_{Z,tri,L_y,d} = V_{Z,tri,L_y,d} H_{Z,tri,L_y,d} \cr\cr T_{Z,hc,L_y,d} &
= & (V_{Z,hc,L_y,d} H_{Z,hc,L_y,d,2}) (V_{Z,hc,L_y,d} H_{Z,hc,L_y,d,1})
\equiv T_{Z,hc,L_y,d,2} T_{Z,hc,L_y,d,1} \ . 
\label{transfermatrix}
\eeqs

As discussed above, the definition of $b^{(d)}$ does not include the possible
permutations of black circles. In principle, there can be $d!$ eigenvalues with
their coefficients $b_j^{(d)}$ added up to $b^{(d)}$. In our explicit
calculation for toroidal strips, $b^{(d)}$ decomposes into two $b_j^{(d)}$
(with possible multiplication by an even integer) for the square and honeycomb lattices when $1 <
d < L_y$ up to $L_y=4$ and for the triangular lattice when $1 < d \le L_y$ up
to $L_y=3$. As discussed above, we know that there is only one coefficient for
both $d=0$ and $d=1$, i.e., $b^{(0)}=1$, $b^{(1)}=q-1$. For $d=2,3$, the
coefficients are
\beqs
b_1^{(2)} & = & \frac12 q(q-3) \cr\cr
b_2^{(2)} & = & \frac12 (q-1)(q-2) \cr\cr
b_1^{(3)} & = & \frac13 (q-1)(q^2-5q+3) \cr\cr 
& = & \frac16 q(q-1)(q-5) + \frac16 (q-1)(q-2)(q-3) \cr\cr
b_2^{(3)} & = & \frac16 q(q-2)(q-4)
\label{bdi}
\eeqs
with possible multiplication. In contrast to cyclic and M\"obius strips,
certain eigenvalues for toroidal strips may appear in more than one level, so
that their coefficients are the summation of the corresponding $b_j^{(d)}$'s.

For the Klein bottle strip of the square and triangular lattices or the
honeycomb lattice with $L_y$ even, the sum of coefficients is the same as the
sum for the corresponding strip with M\"obius longitudinal boundary conditions,
\beq
C_{Z,L_y,Kb} \equiv \sum_{j=1}^{N_{Z,L_y,Kb,\lambda}} c_{Z,L_y,Kb,j} = \cases{ q^{L_y/2} & for even $L_y$ \cr
q^{(L_y+1)/2} & for odd $L_y$ \cr } \ .
\label{ctsumkb}
\eeq
Consider the coefficients for Klein bottle strips. For the square and honeycomb
lattices, when the longitudinal boundary condition is changed from toroidal to
Klein bottle, we observe the following changes of coefficients:
\beqs
b^{(0)} & \to & \pm b^{(0)} \cr\cr
b^{(1)} & \to & \pm b^{(1)} \cr\cr
b_1^{(2)} & \to & \pm b_1^{(2)} \cr\cr
b_2^{(2)} & \to & \pm b_2^{(2)} \cr\cr
b_1^{(3)} & \to & \pm b^{(1)} \ .
\label{bchange}
\eeqs
We find that certain coefficients become zero when the boundary condition is changed from toroidal to Klein bottle, and therefore the number of eigenvalues for Klein bottle strips always appears
to be less than the number for the corresponding toroidal strips. This has been
observed before in Ref. \cite{tk,tor4}. The partition functions for Klein
bottle strips have the same expressions as eqs. (\ref{zgsumtor}) and
(\ref{tgsumtor}) with $b_j^{(d)}$ replaced by appropriate coefficients, as in
eq. (\ref{bchange}). Illustrative calculations will be given below.

\section{Zero-temperature Potts antiferromagnet}

For the chromatic polynomials, it is necessary that adjacent vertices are not
assigned the same color, i.e., in the partition diagrams, adjacent vertices
cannot be connected by color-connection lines. For the square and triangular
lattices, a transverse slice is a circuit graph; while for the honeycomb
lattice, a transverse slice is a set of $L_y/2$ two-vertex tree graphs.

\subsection{Square and triangular lattices}

For the square and triangular lattices, the second partition of the degree $d=0$ subspace and the third partition of the degree $d=1$
subspace for width $L_y=2$ in Fig. \ref{L2partitions} are not allowed. For
$L_y=3$ in Fig. \ref{L3partitions}, we keep the first partition in the $d=0$
subspace, the first three partitions of the $d=1$ subspace, and the first 
six partitions of the $d=2$ subspace. Let us denote the
reduced size of the transfer matrix for the square and triangular lattices as
$n_{P,tor}(L_y,d)$, excluding permutations of black circles. The
sum of all coefficients is the dimension of the total transfer matrix
(independent of $L_y$), which is the chromatic polynomial of the circuit graph
given by Theorem 10 of \cite{cf}, so that for $L_y \ge 2$, 
\beq
C_{P,\Lambda,L_y} = \sum _{d=0}^{L_y} n_{P,tor}(L_y,d)b^{(d)} = (q-1)^{L_y} + (q-1)(-1)^{L_y} \qquad \mbox{for} \ \Lambda=sq, tri \ .
\label{cpsumtor}
\eeq
By substituting the expression for $b^{(d)}$ in eq. (\ref{czsumtor}), we
determine the $n_{P,tor}(L_y,d)$. We list the first few numbers
$n_{P,tor}(L_y,d)$ in Table \ref{nptabletor}. The strip with $L_y=1$ is obtained by identifying the free boundaries of the cyclic strip with $L_y=2$, namely, each vertex is attached by a loop such that its chromatic polynomial is identically zero.  The column $n_{P,tor}(L_y,0)$ in
Table \ref{nptabletor} is the number of non-crossing non-nearest-neighbor
partitions of $L_y$ vertices on a circle, and is given by the Riordan number
$R_{L_y}$ \cite{rpstanley, bernhart99}. If we compare $n_{P,tor}(L_y,0)$ with
the corresponding number $n_{P,cyc}(L_y,0)$ (which was denoted as $n_P(L_y,d)$)
for cyclic strips, given in Table 1 of \cite{cf}, we find the relation
\beq
n_{P,cyc}(L_y,0) - n_{P,tor}(L_y,0) = n_{P,tor}(L_y-1,0) \qquad \mbox{for} \ L_y \ge 2 \ .
\label{nPd0cyctor}
\eeq
This can be understood as follows.  For toroidal strips we have additional
bonds connecting the top and bottom vertices of the corresponding cyclic
strips, so these vertices cannot have the same color. That is, the $d=0$
partitions for cyclic strips with the top and bottom vertices connected (having
the same color) is not allowed for the corresponding toroidal strips. The
number of these partitions is given by $n_{P,tor}(L_y-1,0)$. This is clear, 
since we know $n_{P,cyc}(L_y,0) = M_{L_y-1}$ \cite{cf}, where $M_n$ is the
Motzkin number, and the relation $M_n = R_n + R_{n+1}$. We notice the following
recursion relations:
\beqs
n_{P,tor}(L_y,1) & = & n_{P,tor}(L_y,0) + n_{P,tor}(L_y-1,1) + n_{P,tor}(L_y-1,2) + (-1)^{L_y} \qquad \mbox{for} \ L_y \ge 2 \cr\cr
n_{P,tor}(L_y,2) & = & -n_{P,tor}(L_y-1,0) + 2n_{P,tor}(L_y-1,1) + n_{P,tor}(L_y-1,2) \cr\cr
& & + n_{P,tor}(L_y-1,3) + (-1)^{L_y} \qquad \mbox{for} \ L_y \ge 2 \cr\cr
n_{P,tor}(L_y,d) & = & n_{P,tor}(L_y-1,d-1) + n_{P,tor}(L_y-1,d) + n_{P,tor}(L_y-1,d+1) \cr\cr
& & \qquad \mbox{for} \ L_y \ge 3, \ 3 \le d \le L_y  \ .
\eeqs
The last line here is the same as the recursion relation for $n_{P,cyc}(L_y,d)$
when $1 \le d \le L_y$ for cyclic strips \cite{cf}. By substituting $q=0$ into
eq. (\ref{cpsumtor}) and using eq. (\ref{bdq0}), we have
\beq
\sum _{d=0}^{L_y} n_{P,tor}(L_y,d) (-1)^{d} = 0 \qquad {\rm for} \ \ \Lambda=sq,tri \ . 
\eeq

\begin{table}[htbp]
\caption{\footnotesize{Table of numbers $n_{P,tor}(L_y,d)$ for the square and triangular lattices.}}
\begin{center}
\begin{tabular}{|c|c|c|c|c|c|c|c|c|c|c|c|}
\hline\hline $L_y \backslash \ d$ 
   & 0   & 1   & 2    & 3   & 4   & 5   & 6  & 7  & 8 & 9 & 10 \\ \hline\hline
1  & 0   & 0   &      &     &     &     &    &    &   &   &    \\ \hline
2  & 1   & 2   & 1    &     &     &     &    &    &   &   &    \\ \hline
3  & 1   & 3   & 3    & 1   &     &     &    &    &   &   &    \\ \hline
4  & 3   & 10  & 10   & 4   & 1   &     &    &    &   &   &    \\ \hline
5  & 6   & 25  & 30   & 15  & 5   &  1  &    &    &   &   &    \\ \hline
6  & 15  & 71  & 90   & 50  & 21  &  6  & 1  &    &   &   &    \\ \hline
7  & 36  & 196 & 266  & 161 & 77  &  28 & 7  &  1 &   &   &    \\ \hline
8  & 91  & 554 & 784  & 504 & 266 & 112 & 36 &  8 & 1 &   &    \\ \hline
9  & 232 & 1569& 2304 &1554 & 882 & 414 & 156& 45 & 9 & 1 &    \\ \hline
10 & 603 & 4477& 6765 &4740 &2850 &1452 &615 &210 &55 &10 & 1  \\ \hline\hline
\end{tabular}
\end{center}
\label{nptabletor}
\end{table}

Let us denote the number of diagonal dissections of a convex $n$-gon into $k$
regions as $D(n,k)$, which is sequence A033282 in \cite{sl}. (See also
Ref. \cite{Flajolet99}.)
\beq
D(n,k) = \frac{1}{k} {n-3 \choose k-1} {n+k-2 \choose k-1} \qquad \mbox{for} \ n \ge 3, 1 \le k \le n-2 \ .
\eeq
The first few numbers are $D(3,1)=1$, $D(4,1)=1$, $D(4,2)=2$, $D(5,1)=1$,
$D(5,2)=5$, $D(5,3)=5$. We also know that the Riordan number for $n > 1$ is
given by the number of dissections of a convex polygon by nonintersecting
diagonals with $n+1$ edges. To relate $D(n,k)$ with the Riordan number, define
$D^\prime (n,k)$ as 
\beq
D^\prime (n,k) = D(n+2-k,k) = \frac{1}{k} {n-1-k \choose k-1} {n \choose k-1} \qquad \mbox{for} \ n \ge 2 \ .
\eeq
It is clear that the entries of $D^\prime (n,k)$ are zero for $[n/2]+1 \le k
\le n$. Let us also define $D^\prime (1,1) \equiv 0=n_{P,tor}(1,0)$, and list the first few rows of the triangle of $D^\prime (n,k)$ for $n \ge 1, 1 \le k \le n$ in Fig. \ref{Dprimetriangle}.
The summation of each row in the triangle gives the Riordan number:
\beq
\sum _{k=1}^{[L_y/2]} D^\prime (L_y,k) = R_{L_y} = n_{P,tor}(L_y,0) \ .
\eeq
For $d=1$, we need to assign a color for each component of level 0 partitions,
so that
\beq
n_{P,tor}(L_y,1) = \sum_{k=1}^{[L_y/2]} (L_y+1-k)D^\prime (L_y,k) = \sum_{k=1}^{[L_y/2]} {L_y \choose k} {L_y-1-k \choose k-1} \ .
\eeq

\begin{figure}[htbp]
\begin{center}
\begin{tabular}{ccccccccccccccccc}
  &   &   &   &    &    &    &    & 0 &   &   &   &   &   &   &   &   \\
  &   &   &   &    &    &    & 1  &   & 0 &   &   &   &   &   &   &   \\
  &   &   &   &    &    & 1  &    & 0 &   & 0 &   &   &   &   &   &   \\
  &   &   &   &    & 1  &    & 2  &   & 0 &   & 0 &   &   &   &   &   \\
  &   &   &   & 1  &    & 5  &    & 0 &   & 0 &   & 0 &   &   &   &   \\
  &   &   & 1 &    & 9  &    & 5  &   & 0 &   & 0 &   & 0 &   &   &   \\
  &   & 1 &   & 14 &    & 21 &    & 0 &   & 0 &   & 0 &   & 0 &   &   \\
  & 1 &   & 20&    & 56 &    & 14 &   & 0 &   & 0 &   & 0 &   & 0 &   \\
1 &   & 27&   & 120&    & 84 &    & 0 &   & 0 &   & 0 &   & 0 &   & 0
\end{tabular}
\end{center}
\caption{\footnotesize{Triangle of $D^\prime (n,k)$.}}
\label{Dprimetriangle}
\end{figure}

For entries $n_{P,tor}(L_y,d)$ in Table \ref{nptabletor} with $d > 1$, we find
the following relation with known number sequences. Motivated by
eq. (\ref{nPd0cyctor}), let us define $n^\prime _{P,tor}(L_y,d)$ such that
$n^\prime _{P,tor}(1,d) = n_{P,tor}(1,d)$ and
\beq
n^\prime _{P,tor}(L_y,d) = n_{P,tor}(L_y,d)+n_{P,tor}(L_y-1,d) \qquad \mbox{for} \ L_y \ge 2 \ .
\eeq
We list the first few numbers $n^\prime _{P,tor}(L_y,d)$ in Table
\ref{npprimetabletor}.  Comparing these with the triangular array given by
sequence A025564 in \cite{sl} (which result from pairwise sums of entries in
the trinomial array (sequence A027907)), we find that $n^\prime_{P,tor}(L_y,d)$
for $d \ge 2$ are columns in the array. The first few rows of this triangular
array are listed in Fig. \ref{A025564triangle}. Let us denote a particular
entry as $T(n,k)$; then $T(n,k)=T(n-1,k-2)+T(n-1,k-1)+T(n-1,k)$ starting with
rows $[1]$,$[1,2,1]$,$[1,3,4,3,1]$. We find the relation
\beq
n^\prime_{P,tor}(L_y,d) = T(L_y+1,L_y+1-d) = T(L_y+1,L_y+1+d) \qquad \mbox{for} \ d \ge 2 \ .
\eeq

We list the first few $n^\prime_{P,tor}(L_y,d)-n_{P,cyc}(L_y,d)$ for $L_y \ge
2$ in Table \ref{npprimetabletorcyc}. Because of eq. (\ref{nPd0cyctor}), we
have $n^\prime_{P,tor}(L_y,0) - n_{P,cyc}(L_y,0) =0$. For $d=1$, we find
$n^\prime_{P,tor}(L_y,1) - n_{P,cyc}(L_y,1)$ is given by sequence A005775 in
\cite{sl}, the elements of which are the numbers of compact rooted directed
animals of size $L_y \ge 3$ having 3 source points. Combining the results in 
Table \ref{npprimetabletorcyc} and Table \ref{npprimetabletor}, we infer 
the relation
\beq
n^\prime_{P,tor}(L_y,d) - n_{P,cyc}(L_y,d) = n^\prime_{P,tor}(L_y,d+1) \qquad \mbox{for} \ d \ge 2 \ ,
\eeq
or equivalently,
\beq
n^\prime_{P,tor}(L_y,d) = \sum_{d^\prime=d}^{L_y} n_{P,cyc}(L_y,d^\prime) \qquad \mbox{for} \ d \ge 2 \ .
\eeq

We recall that the sum of the coefficients in (\ref{cpsum}) for a Klein bottle
strip of the square or triangular lattice of width $L_y$ is (independent of
$L_x$), given as Theorem 11 in \cite{cf}, is 
\beq
C_{P,L_y,Kb} = \sum_{j=1}^{N_{P,L_y,Kb,\lambda}} c_{P,L_y,Kb,j} = 0 
\label{cpsumklein}
\eeq

\begin{table}[htbp]
\caption{\footnotesize{Table of numbers $n^\prime_{P,tor}(L_y,d)$.}}
\begin{center}
\begin{tabular}{|c|c|c|c|c|c|c|c|c|c|c|c|}
\hline\hline $L_y \backslash \ d$ 
   & 0   & 1   & 2    & 3   & 4   & 5   & 6  & 7  & 8 & 9 & 10 \\ \hline\hline
1  & 0   & 0   &      &     &     &     &    &    &   &   &    \\ \hline
2  & 1   & 2   & 1    &     &     &     &    &    &   &   &    \\ \hline
3  & 2   & 5   & 4    & 1   &     &     &    &    &   &   &    \\ \hline
4  & 4   & 13  & 13   & 5   & 1   &     &    &    &   &   &    \\ \hline
5  & 9   & 35  & 40   & 19  & 6   &  1  &    &    &   &   &    \\ \hline
6  & 21  & 96  & 120  & 65  & 26  &  7  & 1  &    &   &   &    \\ \hline
7  & 51  & 267 & 356  & 211 & 98  &  34 & 8  &  1 &   &   &    \\ \hline
8  & 127 & 750 & 1050 & 665 & 343 & 140 & 43 &  9 & 1 &   &    \\ \hline
9  & 323 & 2123& 3088 &2058 &1148 & 526 & 192& 53 &10 & 1 &    \\ \hline
10 & 835 & 6046& 9069 &6294 &3732 &1866 &771 &255 &64 &11 & 1  \\ \hline\hline 
\end{tabular}
\end{center}
\label{npprimetabletor}
\end{table}

\begin{figure}[htbp]
\begin{center}
\begin{tabular}{ccccccccccccccccc}
  &   &   &   &   &   &    &    &  1 &    &    &   &   &   &   &   &   \\
  &   &   &   &   &   &    & 1  &  2 &  1 &    &   &   &   &   &   &   \\
  &   &   &   &   &   & 1  & 3  &  4 &  3 &  1 &   &   &   &   &   &   \\
  &   &   &   &   & 1 & 4  & 8  &  10&  8 &  4 & 1 &   &   &   &   &   \\
  &   &   &   & 1 & 5 & 13 & 22 &  26&  22&  13& 5 & 1 &   &   &   &   \\
  &   &   & 1 & 6 & 19& 40 & 61 &  70&  61&  40& 19& 6 & 1 &   &   &   \\
  &   & 1 & 7 & 26& 65& 120& 171& 192& 171& 120& 65& 26& 7 & 1 &   &   \\
  & 1 & 8 & 34& 98&211& 356& 483& 534& 483& 356&211& 98& 34& 8 & 1 &   \\
1 & 9 & 43&140&343&665&1050&1373&1500&1373&1050&665&343&140& 43& 9 & 1
\end{tabular}
\end{center}
\caption{\footnotesize{Triangle of $T(n,k)$.}}
\label{A025564triangle}
\end{figure}

\begin{table}[htbp]
\caption{\footnotesize{Table of numbers $n^\prime_{P,tor}(L_y,d)-n_{P,cyc}(L_y,d)$.}}
\begin{center}
\begin{tabular}{|c|c|c|c|c|c|c|c|c|c|c|c|}
\hline\hline $L_y \backslash \ d$ 
   & 0 & 1   & 2   & 3   & 4   & 5  & 6  & 7 & 8 & 9 & 10\\ \hline\hline
2  & 0 & 0   & 0   &     &     &    &    &   &   &   &   \\ \hline
3  & 0 & 1   & 1   & 0   &     &    &    &   &   &   &   \\ \hline
4  & 0 & 4   & 5   & 1   &  0  &    &    &   &   &   &   \\ \hline
5  & 0 & 14  & 19  & 6   &  1  & 0  &    &   &   &   &   \\ \hline
6  & 0 & 45  & 65  & 26  &  7  & 1  &  0 &   &   &   &   \\ \hline
7  & 0 & 140 & 211 & 98  &  34 & 8  &  1 & 0 &   &   &   \\ \hline
8  & 0 & 427 & 665 & 343 & 140 & 43 &  9 & 1 & 0 &   &   \\ \hline
9  & 0 & 1288& 2058&1148 & 526 & 192& 53 &10 & 1 & 0 &   \\ \hline
10 & 0 & 3858& 6294&3732 &1866 & 771&255 &64 &11 & 1 & 0 \\ \hline\hline 
\end{tabular}
\end{center}
\label{npprimetabletorcyc}
\end{table}

\subsection{Honeycomb lattice}

For a strip of the honeycomb lattice with toroidal boundary conditions, the
width $L_y$ must be even. In each transverse slice of such a strip, the $L_y$
vertices are connected in a pairwise manner, just as for the strips with cyclic
boundary conditions. Let us denote the reduced size of the transfer matrix as
$n_{P,tor,hc}(L_y,d)$, without considering permutations of black circles 
in the
partition diagram. The sum of all coefficients is the dimension of the total
transfer matrix (independent of $L_x$), which is the same as eq. (6.19) of
\cite{hca} for cyclic strips, namely 
\beq
C_{P,hc,L_y} = \sum _{d=0}^{L_y} n_{P,tor,hc}(L_y,d)b^{(d)} = \left( q(q-1) \right)^{L_y/2} \ .
\label{cpsumtorhc}
\eeq
By substituting the expression for $b^{(d)}$ in eq. (\ref{czsumtor}), we 
determine the $n_{P,tor,hc}(L_y,d)$. We list the first few numbers
$n_{P,tor,hc}(L_y,d)$ in Table \ref{nptabletorhc}. We infer the following
recursion relations for $L_y \ge 4$:
\beqs
n_{P,tor,hc}(L_y,0) & = & -n_{P,tor,hc}(L_y-2,0) + 4n_{P,tor,hc}(L_y-2,1) - n_{P,tor,hc}(L_y-2,2) \cr\cr
& & - n_{P,tor,hc}(L_y-2,3) \cr\cr
n_{P,tor,hc}(L_y,1) & = & -3n_{P,tor,hc}(L_y-2,0) + 10n_{P,tor,hc}(L_y-2,1) + n_{P,tor,hc}(L_y-2,2) \cr\cr
n_{P,tor,hc}(L_y,2) & = & -3n_{P,tor,hc}(L_y-2,0) + 8n_{P,tor,hc}(L_y-2,1) + 4n_{P,tor,hc}(L_y-2,2) \cr\cr
& & + 3n_{P,tor,hc}(L_y-2,3) + n_{P,tor,hc}(L_y-2,4) \cr\cr
n_{P,tor,hc}(L_y,3) & = & -n_{P,tor,hc}(L_y-2,0) + 2n_{P,tor,hc}(L_y-2,1) + 3n_{P,tor,hc}(L_y-2,2) \cr\cr
& & + 4n_{P,tor,hc}(L_y-2,3) + 3n_{P,tor,hc}(L_y-2,4) + n_{P,tor,hc}(L_y-2,5) \cr\cr
n_{P,tor,hc}(L_y,d) & = & n_{P,tor,hc}(L_y-2,d-2) + 3n_{P,tor,hc}(L_y-2,d-1) + 4n_{P,tor,hc}(L_y-2,d) \cr\cr
& & + 3n_{P,tor,hc}(L_y-2,d+1) + n_{P,tor,hc}(L_y-2,d+2) \quad \mbox{for} \ 4 \le d \le L_y  \ . \qquad
\eeqs
By substituting $q=0$ into eq. (\ref{cpsumtorhc}) and using eq. (\ref{bdq0}),
we have
\beq
\sum _{d=0}^{L_y} n_{P,tor,hc}(L_y,d) (-1)^{d} = 0 \ . 
\eeq

\begin{table}[htbp]
\caption{\footnotesize{Table of numbers $n_{P,tor,hc}(L_y,d)$ for the honeycomb lattice.}}
\begin{center}
\begin{tabular}{|c|c|c|c|c|c|c|c|c|c|c|c|}
\hline\hline $L_y \backslash \ d$ 
   & 0   & 1   & 2    & 3   & 4   & 5   & 6  & 7  & 8 & 9 & 10 \\ \hline\hline
2  & 1   & 2   & 1    &     &     &     &    &    &   &   &    \\ \hline
4  & 6   & 18  & 17   & 6   & 1   &     &    &    &   &   &    \\ \hline
6  & 43  & 179 & 213  & 108 & 39  &  9  & 1  &    &   &   &    \\ \hline
8  & 352 & 1874& 2518 &1512 & 721 & 264 & 70 & 12 & 1 &   &    \\ \hline
10 &3114 &20202&29265 &19425&10800&4953 &1830&525 &110&15 & 1  \\ \hline\hline
\end{tabular}
\end{center}
\label{nptabletorhc}
\end{table}

It is clear that the number of degree $d=0$ partitions $n_{P,tor,hc}(L_y,0)$ is
the same as the corresponding number $n_{P,cyc,hc}(L_y,0)$ for the
corresponding cyclic strips (which was denoted as $n_P(hc,L_y,d)$ in
\cite{hca}). Combining these results with those in Table 2 of \cite{hca}, we 
observe that 
\beq
n_{P,tor,hc}(L_y,1) = \frac{N_{P,cyc,hc,L_y,\lambda}}{2}
\eeq
where $N_{P,cyc,hc,L_y,\lambda}$ is the total number of $\lambda$ for 
cyclic
strips. We list the first few $n_{P,tor,hc}(L_y,d)-n_{P,cyc,hc}(L_y,d)$ in
Table \ref{nptabletorhccyc}. Compare Table \ref{nptabletorhccyc} with Table
\ref{nptabletorhc}, it is easy to see the relation
\beq
n_{P,tor,hc}(L_y,d) - n_{P,cyc,hc}(L_y,d) = n_{P,tor,hc}(L_y,d+1) \qquad \mbox{for} \ d \ge 2 \ ,
\label{nztorcychc}
\eeq
or equivalently,
\beq
n_{P,tor,hc}(L_y,d) = \sum_{d^\prime=d}^{L_y} n_{P,cyc,hc}(L_y,d^\prime) \qquad \mbox{for} \ d \ge 2 \ .
\eeq
We also have
\beq
n_{P,tor,hc}(L_y,1) - n_{P,cyc,hc}(L_y,1) = \sum_{d^\prime=3}^{L_y} (-1)^{d^\prime+1} n_{P,tor,hc}(L_y,d^\prime) = \sum_{{\rm odd} \ d^\prime=3}^{L_y-1} n_{P,cyc,hc}(L_y,d^\prime) \ .
\eeq

\begin{table}[htbp]
\caption{\footnotesize{Table of numbers $n_{P,tor,hc}(L_y,d)-n_{P,cyc,hc}(L_y,d)$ for the honeycomb lattice.}}
\begin{center}
\begin{tabular}{|c|c|c|c|c|c|c|c|c|c|c|c|}
\hline\hline $L_y \backslash \ d$ 
   & 0   & 1   & 2    & 3   & 4   & 5   & 6  & 7  & 8 & 9 & 10 \\ \hline\hline
2  & 0   & 0   & 0    &     &     &     &    &    &   &   &    \\ \hline
4  & 0   & 5   & 6    & 1   & 0   &     &    &    &   &   &    \\ \hline
6  & 0   & 77  & 108  & 39  & 9   &  1  & 0  &    &   &   &    \\ \hline
8  & 0   & 996 & 1512 & 721 & 264 & 70  & 12 & 1  & 0 &   &    \\ \hline
10 & 0   &12177& 19425&10800&4953 &1830 &525 &110 &15 & 1 & 0  \\ \hline\hline
\end{tabular}
\end{center}
\label{nptabletorhccyc}
\end{table}

\section{Properties of Transfer Matrices at Special Values of Parameters}

In this section we derive some properties of the transfer matrices 
$T_{Z,\Lambda,L_y,d}$ at special values of $q$ and $v$, and, correspondingly,
properties of $T_{T,\Lambda,L_y,d}$ at special values of $x$ and $y$.  

\subsection{ $v=0$ }

 From (\ref{zfun}) or (\ref{cluster}) it follows that for any graph $G$, the
Potts model partition function $Z(G,q,v)$ satisfies 
\beq
Z(G,q,0)=q^{n(G)} \ . 
\label{zv0}
\eeq
Since this holds for arbitrary values of $q$, in the context of the lattice 
strips considered here, it implies that 
\beq
(T_{Z,\Lambda,L_y,d})_{v=0} = 0 \quad {\rm for} \quad 1 \le d \le L_y \ , 
\label{tzlydv0}
\eeq
i.e. these are zero matrices. Secondly, restricting to toroidal strips for
simplicity, and using the basic results $n=L_yL_x=L_y m$ for $\Lambda=sq,tri$
and $n=2L_y m$ for $\Lambda=hc$, eq. (\ref{zv0}) implies that
\beq
Tr[(T_{Z,\Lambda,L_y})^m]_{v=0} 
=\cases{q^{L_y m}&\ for \ $\Lambda=sq,tri$ \cr
             q^{2L_y m} &\ for \ $\Lambda=hc$  } \ .
\label{TThc1xl} 
\eeq
With our explicit calculations, we have 
\beq
(T_{Z,\Lambda,L_y,0})_{jk} = 0 \quad {\rm for} \ \ v=0, \quad {\rm and} \quad
j \ge 2 \ , 
\label{Tjkv0}
\eeq
i.e., all rows of these matrices except the first vanish, and
\beq
(T_{Z,\Lambda,L_y,0})_{11}=q^{p L_y} \quad {\rm for} \quad v=0 \ , 
\label{T11v0}
\eeq
where $p$ was given in eq. (\ref{powerp}). 
For this $v=0$ case, since all of the rows except the first are zero, the
elements $(T_{Z,\Lambda,L_y,0})_{1k}$ for $k \ge 2$ do not enter into
$Tr[(T_{Z,\Lambda,L_y,0})^m]$, which just reduces to the $m$'th power of the
(1,1) element:
\beq
Tr[(T_{Z,\Lambda,L_y,0})^m]=[(T_{Z,\Lambda,L_y,0})_{11}]^m \ . 
\label{tracev0}
\eeq
Corresponding to this, all of the eigenvalues $\lambda_{Z,\Lambda,L_y,d,j}$
vanish except for one, which is equal to $(T_{Z,\Lambda,L_y,0})_{11}$ in
eq. (\ref{T11v0}).  As will be seen, this is reflected in the property that
$det(T_{Z,\Lambda,L_y,d})$ has a nonzero power of $v$ as a factor for all of
the lattice-$\Lambda$ strips considered here.  Note that the condition $v=0$ is
equivalent to the Tutte variable condition $y=1$.

\subsection{ $q=0$ } 

Another fundamental relation that follows, e.g., by setting $q=0$ in eq. 
(\ref{cluster}), is 
\beq
Z(G,0,v)=0 \ . 
\label{zq0}
\eeq
Now the coefficients $b^{(d)}$ evaluated at $q=0$ satisfy
eq.(\ref{bdq0}). Although $b^{(d)}$ decomposes into two $b^{(d)}_j$ for $d >
1$, one of them is equal to zero and the other is $(-1)^d$ at $q=0$ for the
cases we have, namely $b^{(2)}_1=0$, $b^{(2)}_2=1$, $b^{(3)}_1=-1$,
$b^{(3)}_2=0$. Hence, in terms of transfer matrices, we derive the sum rule
\beq
\sum_{0 \le {\rm even} \ d \le L_y} \sum_j (\lambda_{Z,\Lambda,L_y,d,j}(0,v))^m - \sum_{1 \le {\rm odd} \ d \le L_y} \sum_j (\lambda_{Z,\Lambda,L_y,d,j}(0,v))^m =0 \ .
\label{zq0sumrule}
\eeq
This is similar to a sum rule that we obtained in Ref. \cite{s5}.  Since it
applies for arbitrary $m$, it implies that there must be a pairwise
cancellation between various eigenvalues in different degree-$d$ subspaces,
which, in turn, implies that at $q=0$ there are equalities between these
eigenvalues.  For example, for the $L_y=2$ strip of the square lattice and
$q=0$, two of the eigenvalues of $T_{Z,sq,2,1}$ become equal to the eigenvalues
of $T_{Z,sq,2,0}$, while the third eigenvalue of $T_{Z,sq,2,1}$ becomes equal
to $\lambda_{Z,sq,2,2}=v^2$.  Note that setting $q=0$ does not, in general,
lead to any additional vanishing eigenvalues for the $T_{Z,\Lambda,L_y,d}$ of
the $\Lambda=sq,hc$ lattices, and hence our formulas below for
$T_{Z,\Lambda,L_y,d}$ do not contain overall factors of $q$. However, for the
triangular lattice, certain eigenvalues do vanish at $q=0$.

In terms of the variables $x$ and $y$ in the Tutte polynomial, the value $q=0$
is equivalent to the value $x=1$ (unless $v=0$).  In contrast to the vanishing
of $Z(G,q,v)$ at $q=0$, the Tutte polynomial $T(G,1,y)$ is nonzero for general
$y$.  This different behavior can be traced to the feature that in
eq. (\ref{zt}), $Z(G,q,v)$ is proportional to $T(G,x,y)$ multiplied by the
factor $(x-1)^{k(G)}=(x-1)$, so at $x=1$, $Z(G,0,v)=0$ even if $T(G,1,y) \ne
0$.

\subsection{$v=-1$, i.e., $y=0$} 

As discussed above, the special value $v=-1$, i.e., $y=0$, corresponds to the
zero-temperature Potts antiferromagnet, and in this case the Potts model
partition function reduces to the chromatic polynomial, as indicated in
eq. (\ref{zp}).  In the previous two sections we have determined how the
dimensions $n_{Z,tor}(L_y,d)$ for the square, triangular, and honeycomb lattice
strips reduce to the dimensions $n_{P,tor}(L_y,d)$.  In all cases, in each
degree-$d$ subspace for $0 \le d \le L_y-1$ for $\Lambda=sq,tri,hc$, some
eigenvalues vanish, so that $n_{P,tor,\Lambda}(L_y,d) <
n_{Z,tor,\Lambda}(L_y,d)$ and hence $det(T_{Z,\Lambda,L_y,d})=0$ at $v=-1$.
This property is reflected in the powers of $(v+1)$ and $y$ that appear,
respectively, in our formulas below for $det(T_{Z,\Lambda,L_y,d})$ and
$det(T_{T,\Lambda,L_y,d})$.

\subsection{$v=-q$, i.e., $x=0$}

For the graph $G=G(V,E)$, setting $x=0$ and $y=1-q$ in the Tutte polynomial
$T(G,x,y)$ yields the flow polynomial $F(G,q)$, which counts the number of
nowhere-0 $q$-flows (without sinks or sources) that there are on $G$
\cite{boll}:
\beq
F(G,q)=(-1)^{e(G)-n(G)+1}T(G,0,1-q) \ . 
\label{ft}
\eeq
Therefore, the flow polynomial for toroidal lattice strips has the form 
\beq
F(\Lambda,L_y \times L_x, tor.,q) = \sum_{d=0}^{L_y} 
\sum_j b^{(d)}_j (\lambda_{F,\Lambda,L_y,d,j})^m \ .
\label{fgsumcyc}
\eeq
We found that for all $L_y \ge 1$ and $0 \le d \le L_y-1$ for $\Lambda=sq,hc$,
when one sets $x=0$ in the Tutte polynomial, or equivalently, $q=-v$ in the
Potts model partition function, some of the eigenvalues in each degree-$d$
subspace vanish, and hence $det(T_{Z,\Lambda,L_y,d})$ and
$det(T_{T,\Lambda,L_y,d})$ vanish.  This is reflected in the powers of $(q+v)$
and $x$ that appear, respectively, in our formulas below for
$det(T_{Z,\Lambda,L_y,d})$ and $det(T_{T,\Lambda,L_y,d})$. The expressions for
the determinants for strips of the triangular lattice are more complicated
because of the fact alluded to above that one has to work with width-$(L_y+1)$
matrices and then identify the two end vertices.

\section{General Results for Toroidal Strips}

In this section we present general results that we have obtained for 
transfer matrices of toroidal strips and their properties. These are valid for
arbitrarily large strip widths $L_y \ge 2$ (as well as arbitrarily great
lengths). 

\subsection{Determinants}

For the square lattice, we find
\beq
det(T_{T,sq,L_y,d}) = (xy)^{d! \ L_y n_{Z,tor}(L_y-1,d)}
\label{detTTsqld}
\eeq
where $n_{Z,tor}(L_y,d)$ was given by eq. (\ref{nzly0})-(\ref{nzlyd}). This
result applies for all $d$, i.e., $0 \le d \le L_y$ since
$n_{Z,tor}(L_y-1,d)=0$ for $d > L_y-1$ (c.f. eq. (\ref{nzup})).  This is
equivalent to the somewhat more complicated expression for
$det(T_{Z,sq,L_y,d})$:
\beq
det(T_{Z,sq,L_y,d}) = 
v^{d! \ L_y n_{Z,tor}(L_y,d)} \biggl [ \Bigl ( 1+\frac{q}{v} \Bigr )
(1+v) \biggr ]^{d! \ L_y n_{Z,tor}(L_y-1,d)} \ . 
\label{detTZsqld}
\eeq
These determinant formulas can be explained as follows. By the arrangement of
the partitions, as shown in Figs. \ref{L2partitions} and \ref{L3partitions},
the matrix $J_{L_y,d,i,i+1}$ has the lower triangular form and the matrix
$D_{L_y,d,i}$ has the upper triangular form. From the definition of the
transfer matrix in eq.  (\ref{transfermatrix}), the determinant of
$T_{Z,sq,L_y,d}$ is the product of the diagonal elements of $H_{Z,sq,L_y,d}$
and $V_{Z,sq,L_y,d}$. The diagonal elements of $H_{Z,sq,L_y,d}$ have the form
$(1+v)^r$, where $r$ is the number of edges in the corresponding partition, and
the diagonal elements of $V_{Z,sq,L_y,d}$ have the form $v^{L_y} (1+q/v)^s$,
where $s$ is the number of vertices which do not connect to any other vertex
in the corresponding partition. Therefore, the power of $(1+v)$ in
eq. (\ref{detTZsqld}) is the sum of the number of nearest-neighbor edges of all
the $(L_y,d)$-partitions. Let us compare the $(L_y,d)$-partitions and
$(L_y-1,d)$ partitions. Since each edge of a $(L_y,d)$-partition corresponds to
adding a new vertex in a $(L_y-1,d)$-partition (at $L_y$ possible places) and
connecting it with the vertex above it, the power of $(1+v)$ is $d! \ L_y
n_{Z,tor}(L_y-1,d)$, where the factor $d!$ comes from the permutation of black
circles. It is clear that the power of $v$ is $d! \ L_y n_{Z,tor}(L_y,d)$. The
power of $(1+q/v)$ in eq. (\ref{detTZsqld}) is the sum of the number of
unconnected vertices of all the $(L_y,d)$-partitions. Now consider the
$(L_y,d)$-partitions and $(L_y-1,d)$ partitions again. Since each unconnected
vertex of a $(L_y,d)$-partition corresponds to adding a unconnected vertex to a
$(L_y-1,d)$-partition in $L_y$ possible ways, the power of $(1+q/v)$ is again $d!
\ L_y n_{Z,tor}(L_y-1,d)$.

Next, taking into account that the generalized multiplicity is $b^{(d)}$, we
have, for the total determinant
\beqs
det(T_{Z,sq,L_y}) 
& = & \prod_{d=0}^{L_y}
    (y-1)^{L_y n_{Z,tor}(L_y,d)b^{(d)}} 
    (x y)^{L_y n_{Z,tor}(L_y-1,d)b^{(d)}} \cr\cr
& = & (y-1)^{L_y \sum_{d=0}^{L_y} n_{Z,tor}(L_y,d)b^{(d)}}
  (x y)^{L_y \sum_{d=0}^{L_y} n_{Z,tor}(L_y-1,d)b^{(d)}} \ . 
\label{detTZsqLaux}
\eeqs
Using eq. (\ref{czsumtor}) together with eq. (\ref{nzup}) so that
$\sum_{d=0}^{L_y} n_{Z,tor}(L_y-1,d)b^{(d)}$=$\sum_{d=0}^{L_y-1} n_{Z,tor}(L_y-1,d)b^{(d)}$, we have, finally,
\beq
det(T_{Z,sq,L_y}) = (y-1)^{L_y q^{L_y}} (x y)^{L_y q^{L_y-1}} \ .
\label{detTZsqL}
\eeq
This agrees with the conjecture given as eq. (3.70) of our earlier Ref.
\cite{s3a} for the determinant of the transfer matrix of the toroidal strip of
the square lattice with arbitrary width $L_y$.

For the honeycomb lattice, we find
\beq 
det(T_{T,hc,L_y,d}) = (x^2 y)^{d! \ L_y n_{Z,tor}(L_y-1,d)} \ . 
\label{detTThcld} 
\eeq
Equivalently,
\beq 
det(T_{Z,hc,L_y,d}) = (v^2)^{d! \ L_y n_{Z,tor}(L_y,d)} \biggl [ \Bigl (
1+\frac{q}{v} \Bigr )^2 (1+v) \biggr ]^{d! \ L_y n_{Z,tor}(L_y-1,d)} \ . 
\label{detTZhcld} 
\eeq
This can be understood as follows: by an argument similar to that given before,
the power of $(1+v)$ is the same as for the square lattice case. Comparing
$T_{Z,sq,L_y,d}$ and $T_{Z,hc,L_y,d}$ in eq. (\ref{transfermatrix}), one sees
that $V_{Z,hc,L_y,d} = V_{Z,sq,L_y,d}$ has been multiplied twice for the
honeycomb lattice, so that the powers of $v$ and $(1+q/v)$ become twice of the
corresponding powers for the square lattice. 

Taking into account that the generalized multiplicity is $b^{(d)}$, the total
determinant for the $hc$ lattice is given by
\beq det(T_{T,hc,L_y}) = (x^2y)^{L_y q^{L_y-1}} \ . 
\label{detTThc} 
\eeq
Equivalently,
\beq 
det(T_{Z,hc,L_y}) = (v^2)^{L_y q^{L_y}} \biggl [ \Bigl (
1+\frac{q}{v} \Bigr )^2 (1+v) \biggr ]^{L_y q^{L_y-1}} \ . 
\label{detTZhc} 
\eeq

It is clear that $det(T_{T,hc,L_y,d})$ is related to $det(T_{T,sq,L_y,d})$ by
the replacement $x \to x^2$ (holding $y$ fixed).  This, together with the fact
that $n_{Z,tor}(L_y,d)$ is the same for these lattices means that the total
determinants $det(T_{T,hc,L_y})$ is related to $det(T_{T,sq,L_y})$ by the same
respective replacements.  Correspondingly, $det(T_{Z,hc,L_y,d})$ is related to
$det(T_{Z,sq,L_y,d})$ by the replacements of the respective factors $v$ by
$v^2$ (cf. eq. (\ref{powerp})) and $(1+q/v)$ by $(1 + q/v)^2$.

\subsection{Traces}

\begin{figure}
\unitlength 1mm \hspace*{5mm}
\begin{picture}(50,30)
\multiput(10,0)(0,6){2}{\circle{1}}
\multiput(10,18)(0,6){3}{\circle{1}}
\multiput(16,0)(0,6){2}{\circle{1}}
\multiput(16,18)(0,6){3}{\circle{1}}
\multiput(10,0)(0,6){2}{\line(1,0){6}}
\multiput(10,18)(0,6){3}{\line(1,0){6}}
\multiput(10,0)(6,0){2}{\line(0,1){6}}
\multiput(10,18)(6,0){2}{\line(0,1){12}}
\put(10,15){\oval(4,30)[l]} 
\put(16,15){\oval(4,30)[r]} 
\multiput(10,12)(6,0){2}{\makebox(0,0){{\small $\vdots$}}}
\multiput(4,0)(18,0){2}{\makebox(0,0){{\footnotesize $L_y$}}}
\put(4,6){\makebox(2,0)[r]{{\footnotesize $L_y-1$}}}
\put(22,6){\makebox(8,0)[r]{{\footnotesize $L_y-1$}}}
\multiput(4,18)(18,0){2}{\makebox(0,0){{\footnotesize $3$}}}
\multiput(4,24)(18,0){2}{\makebox(0,0){{\footnotesize $2$}}}
\multiput(4,30)(18,0){2}{\makebox(0,0){{\footnotesize $1$}}}
\put(36,15){\makebox(0,0){{\small $=$}}}
\multiput(56,0)(0,6){2}{\circle{1}}
\multiput(56,18)(0,6){3}{\circle{1}} 
\put(56,0){\line(0,1){6}} \put(56,18){\line(0,1){12}}
\put(56,15){\oval(4,30)[l]} 
\put(56,12){\makebox(0,0){{\small $\vdots$}}} \multiput(59,0)(0,6){2}{\oval(6,2)}
\multiput(59,18)(0,6){3}{\oval(6,2)}
\put(50,0){\makebox(0,0){{\footnotesize $L_y$}}}
\put(50,6){\makebox(2,0)[r]{{\footnotesize $L_y-1$}}}
\put(50,18){\makebox(0,0){{\footnotesize $3$}}}
\put(50,24){\makebox(0,0){{\footnotesize $2$}}}
\put(50,30){\makebox(0,0){{\footnotesize $1$}}}
\end{picture}

\caption{\footnotesize{$m=1$ graph for the toroidal square lattice.}} \label{sqtrace}
\end{figure}
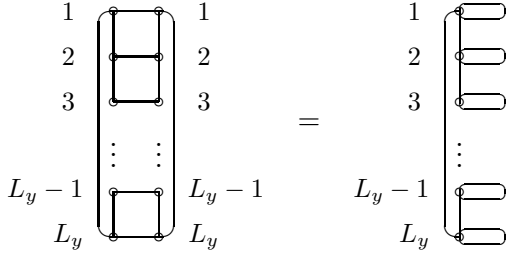

The trace of the total transfer matrix is the $m=1$ case in
eq. (\ref{zgsumtor}) or (\ref{tgsumtor}). For strips of the square lattice,
this corresponds to a $L_y$-vertex circuit with a loop attached to each vertex,
as illustrated in Fig. \ref{sqtrace}. We have $Tr(T_{T,sq,L_y})$ given by the
corresponding Tutte polynomial,
\beq
Tr(T_{T,sq,L_y}) = y^{L_y} \left ( y-1+\frac{x^{L_y}-1}{x-1} \right ) \ . 
\label{traceTTsq}
\eeq
Equivalently, we find 
\beq
Tr(T_{Z,sq,L_y}) = (1+v)^{L_y} [ (v+q)^{L_y}+(q-1)v^{L_y} ] \ . 
\label{traceTZsq}
\eeq
In principle, we can also consider the $m=1$ case for the Klein bottle
strips, but the result is not as simple as that listed here.

\begin{figure}
\unitlength 1mm \hspace*{5mm}
\begin{picture}(50,30)
\multiput(10,0)(0,6){2}{\circle{1}}
\multiput(10,18)(0,6){3}{\circle{1}}
\multiput(16,0)(0,6){2}{\circle{1}}
\multiput(16,18)(0,6){3}{\circle{1}}
\multiput(10,0)(0,6){2}{\line(1,0){6}}
\multiput(10,18)(0,6){3}{\line(1,0){6}}
\multiput(10,0)(6,0){2}{\line(0,1){6}}
\multiput(10,18)(6,0){2}{\line(0,1){12}} \put(10,0){\line(1,1){6}}
\multiput(10,18)(0,6){2}{\line(1,1){6}}
\put(10,30){\line(1,-5){6}}
\put(10,15){\oval(4,30)[l]}
\put(16,15){\oval(4,30)[r]}
\multiput(10,12)(6,0){2}{\makebox(0,0){{\small $\vdots$}}}
\multiput(4,0)(18,0){2}{\makebox(0,0){{\footnotesize $L_y$}}}
\put(4,6){\makebox(2,0)[r]{{\footnotesize $L_y-1$}}}
\put(22,6){\makebox(8,0)[r]{{\footnotesize $L_y-1$}}}
\multiput(4,18)(18,0){2}{\makebox(0,0){{\footnotesize $3$}}}
\multiput(4,24)(18,0){2}{\makebox(0,0){{\footnotesize $2$}}}
\multiput(4,30)(18,0){2}{\makebox(0,0){{\footnotesize $1$}}}
\put(36,15){\makebox(0,0){{\small $=$}}}
\multiput(56,0)(0,6){2}{\circle{1}}
\multiput(56,18)(0,6){3}{\circle{1}} \put(56,3){\oval(2,6)}
\multiput(56,21)(0,6){2}{\oval(2,6)}
\put(56,15){\oval(4,30)[l]}
\put(56,15){\oval(6,30)[l]}
\put(56,12){\makebox(0,0){{\small $\vdots$}}}
\multiput(59,0)(0,6){2}{\oval(6,2)}
\multiput(59,18)(0,6){3}{\oval(6,2)}
\put(50,0){\makebox(0,0){{\footnotesize $L_y$}}}
\put(50,6){\makebox(2,0)[r]{{\footnotesize $L_y-1$}}}
\put(50,18){\makebox(0,0){{\footnotesize $3$}}}
\put(50,24){\makebox(0,0){{\footnotesize $2$}}}
\put(50,30){\makebox(0,0){{\footnotesize $1$}}}
\end{picture}

\caption{\footnotesize{$m=1$ graph for the toroidal triangular
lattice.}} \label{tritrace}
\end{figure}
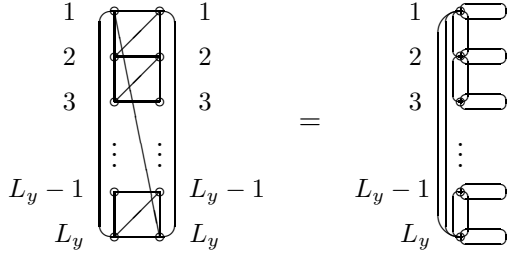

For the triangular lattice, the total trace corresponds to a $L_y$-vertex
circuit with each edge doubled and with a loop attached to each vertex as
illustrated in Fig. \ref{tritrace}. Therefore, the trace is given by the Tutte
polynomial of this graph
\beq
Tr(T_{T,tri,L_y}) = y^{L_y} \left ( (y-1)(y+1)^{L_y} + \sum_{j=0}^{L_y-1} (y+1)^j(x+y)^{L_y-1-j} \right ) \ . 
\label{traceTTtri}
\eeq
Equivalently,
\beq
Tr(T_{Z,tri,L_y}) = q(v+1)^{L_y} \left ( (v^2+2v)^{L_y} + \sum_{j=0}^{L_y-1} (v^2+2v)^j(v^2+2v+q)^{L_y-1-j} \right ) \ . 
\label{traceTZtri}
\eeq

\begin{figure}
\unitlength 1mm \hspace*{5mm}
\begin{picture}(70,36)
\multiput(10,0)(0,6){2}{\circle{1}}
\multiput(10,18)(0,6){4}{\circle{1}}
\multiput(16,0)(0,6){2}{\circle{1}}
\multiput(16,18)(0,6){4}{\circle{1}}
\multiput(22,0)(0,6){2}{\circle{1}}
\multiput(22,18)(0,6){4}{\circle{1}}
\multiput(10,0)(0,6){2}{\line(1,0){12}}
\multiput(10,18)(0,6){4}{\line(1,0){12}}
\multiput(10,0)(12,0){2}{\line(0,1){6}}
\multiput(10,18)(12,0){2}{\line(0,1){6}}
\multiput(10,30)(12,0){2}{\line(0,1){6}}
\put(16,24){\line(0,1){6}} 
\put(16,12){\makebox(0,0){{\small $\vdots$}}}
\put(16,18){\oval(4,36)[r]}
\multiput(4,0)(24,0){2}{\makebox(0,0){{\footnotesize $L_y$}}} \put(4,6){\makebox(2,0)[r]{{\footnotesize $L_y-1$}}}
\put(30,6){\makebox(6,0)[r]{{\footnotesize $L_y-1$}}} \multiput(4,18)(24,0){2}{\makebox(0,0){{\footnotesize $4$}}} \multiput(4,24)(24,0){2}{\makebox(0,0){{\footnotesize $3$}}} \multiput(4,30)(24,0){2}{\makebox(0,0){{\footnotesize $2$}}} \multiput(4,36)(24,0){2}{\makebox(0,0){{\footnotesize $1$}}} \put(42,18){\makebox(0,0){{\small $=$}}}
\multiput(62,0)(0,6){2}{\circle{1}}
\multiput(68,0)(0,6){2}{\circle{1}}
\multiput(62,18)(0,6){4}{\circle{1}}
\multiput(68,18)(0,6){4}{\circle{1}} \put(62,0){\line(0,1){6}}
\put(62,18){\line(0,1){6}} \put(68,24){\line(0,1){6}}
\put(62,30){\line(0,1){6}} \multiput(65,0)(0,6){2}{\oval(6,2)}
\multiput(65,18)(0,6){4}{\oval(6,2)}
\put(68,18){\oval(4,36)[r]}
\put(65,12){\makebox(0,0){{\small $\vdots$}}}
\put(56,0){\makebox(0,0){{\footnotesize $L_y$}}}
\put(56,6){\makebox(2,0)[r]{{\footnotesize $L_y-1$}}}
\put(56,18){\makebox(0,0){{\footnotesize $4$}}}
\put(56,24){\makebox(0,0){{\footnotesize $3$}}}
\put(56,30){\makebox(0,0){{\footnotesize $2$}}}
\put(56,36){\makebox(0,0){{\footnotesize $1$}}}
\end{picture}

\caption{\footnotesize{$m=1$ graph for the toroidal honeycomb
lattice with even $L_y$.}} \label{hctrace}
\end{figure}
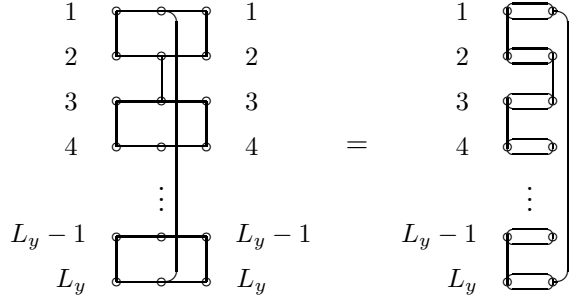

For the honeycomb lattice, the total trace corresponds to a $2L_y$-vertex
circuit with every other edge doubled as shown in
Fig. \ref{hctrace}. Hence, the trace is given by the Tutte polynomial of
this graph
\beq 
Tr(T_{T,hc,L_y}) = (x+y)^{L_y} \left ( \frac{x^{L_y}-1}{x-1} \right ) + (y-1)(y+1)^{L_y} + \sum_{j=0}^{L_y-1} (y+1)^j(x+y)^{L_y-1-j} \ . 
\label{traceTThc} 
\eeq
Equivalently,
\beqs
Tr(T_{Z,hc,L_y}) & = & (v^2+2v+q)^{L_y} \left ( (v+q)^{L_y}-v^{L_y} \right ) \cr\cr & & + qv^{L_y} \left ( (v^2+2v)^{L_y} + \sum_{j=0}^{L_y-1} (v^2+2v)^j(v^2+2v+q)^{L_y-1-j} \right ) \ . 
\label{traceTZhc} 
\eeqs

\subsection{Eigenvalues for $d=L_y$ and $d=L_y-1$ for $\Lambda=sq,hc$}

It was shown earlier \cite{tk,tor4} that the $\lambda$'s for a strip with Klein
bottle boundary conditions are a subset of the $\lambda$'s for the same strip
with torus boundary conditions.  From eq. (\ref{nzlyd}) one knows that there is
only one $\lambda$, denoted as $\lambda_{Z,\Lambda,L_y,L_y}$, for degree
$d=L_y$ for strips of the square and honeycomb lattices because all the
permutation of black circles are equivalent. That is, for this value of $d$,
the transfer matrix reduces to $1 \times 1$, i.e. is a scalar. We found that
for a toroidal or Klein bottle strip of the square, or honeycomb lattice with
width $L_y$,
\beq
\lambda_{T,\Lambda,L_y,L_y}=1 \ . 
\label{lamtutdly}
\eeq
Equivalently, in terms of Potts model variables, 
\beq
\lambda_{Z,sq,L_y,L_y}=v^{L_y} 
\label{lamdlysq}
\eeq
\beq
\lambda_{Z,hc,L_y,L_y}=v^{2L_y} \ . 
\label{lamdlyhc}
\eeq

For the width $L_y$ triangular strips, because one has to start by 
constructing the width-$(L_y+1)$ transfer matrices, there can be more than one
$\lambda$ even for degree $d=L_y$.

Concerning the $\lambda$'s for degree $d=L_y-1$, we find that one of these is
the same, independent of $L_y$, for the toroidal square strips we have
calculated, namely
\beq
\lambda_{T,sq,L_y,L_y-1,1} = x \ .
\label{lamtutsqdm1x}
\eeq
This also appears for degree $d=L_y-1$ in the case of the cyclic square-lattice
strips, and we conjecture that all the toroidal square-lattice strips have this
eigenvalue. For the strips of the square lattice with $L_y=3,4$, there is
another common $\lambda$ for degree $d=L_y-1$, namely
\beq
\lambda_{T,sq,L_y,L_y-1,2} = y \ .
\label{lamtutsqdm1y}
\eeq

For the toroidal honeycomb strip with $L_y=4$, one of the eigenvalues for
degree $d=L_y-1$, say that for $j=1$, is given by
\beq
\lambda_{T,hc,L_y,d=L_y-1,j=1}=x^2
\label{lamtut_hcdm1}
\eeq
as for the corresponding cyclic strip. We conjecture that all of the toroidal
strips of the honeycomb lattice have this eigenvalue. In terms of Potts model
quantities these results are
\beqs
\lambda_{Z,sq,L_y,L_y-1,1} & = & v^{L_y-1}(v+q) \cr\cr
\lambda_{Z,sq,L_y,L_y-1,2} & = & v^{L_y}(v+1) \qquad \mbox{except for} \ L_y=2 \cr\cr
\lambda_{Z,hc,L_y,L_y-1,1} & = & v^{2(L_y-1)}(v+q)^2 \ .
\label{lamsqhcdm1x}
\eeqs
The expressions for the other eigenvalues are, in general, more complicated.

\section{Some Illustrative Calculations}

\subsection{Square-Lattice Strip, $L_{\lowercase{y}}=2$}

The toroidal strip of the square lattice with width $L_y=2$ is equivalent to
the ladder graph with all the transverse edges doubled.  The Potts model
partition function $Z(sq,L_y \times m,q,v)$ and Tutte polynomial $T(sq,L_y
\times m,x,y)$ were calculated for the toroidal and Klein bottle strips of the
square lattice with width $L_y=2$ in Ref. \cite{s3a}.  We express the results
here in terms of transfer matrices $T_{Z,sq,2,d}$. For $d=0$, we have
\beq
T_{Z,sq,2,0} = \left( \begin{array}{cc}
    q^2+4qv+qv^2+5v^2+2v^3 & (q+2v)(1+v)^2 \\
    v^3(2+v)          & v^2(1+v)^2 \end{array} \right ) \ .
\label{TZsq20}
\eeq
The eigenvalues of $T_{Z,sq,2,0}$ are the same as $\lambda_{Z,s2t,(5,6)}$ given
as eqs. (3.16) and (3.17) of \cite{s3a}. For $d=1$, we have
\beq
T_{Z,sq,2,1} = v \left( \begin{array}{ccc}
    q+3v+v^2 & v(2+v) & (1+v)^2 \\
    v(2+v) & q+3v+v^2 & (1+v)^2 \\
    v^2(2+v) & v^2(2+v) & v(1+v)^2 \end{array} \right ) \ .
\label{TZsq21}
\eeq
The eigenvalues of $T_{Z,sq,2,1}$ are the same as $\lambda_{Z,s2t,(2,3,4)}$
given as eqs. (3.13) to (3.15) of \cite{s3a}. The matrix $T_{T,sq,2,2}=v^2$ has
been given above.

\subsection{Square-Lattice Strip, $L_{\lowercase{y}}=3$}

For the $L_y=3$ toroidal and Klein bottle strips of the square lattice, the
Potts model partition function $Z(sq,L_y \times m,q,v)$ and Tutte polynomial
$T(sq,L_y \times m,x,y)$ were calculated in Ref. \cite{s3a}.  We express the
results here in terms of transfer matrices $T_{Z,sq,3,d}$. For $d=0$, we have
\beq
T_{Z,sq,3,0} = \left( \begin{array}{ccccc}
    s_1 & v_1s_2 & v_1s_2 & v_1s_2 & v_1^3s_5 \\
    v^3s_3 & v^2v_1s_4 & v^3v_1v_2 & v^3v_1v_2 & v^2v_1^3 \\
    v^3s_3 & v^3v_1v_2 & v^2v_1s_4 & v^3v_1v_2 & v^2v_1^3 \\
    v^3s_3 & v^3v_1v_2 & v^3v_1v_2 & v^2v_1s_4 & v^2v_1^3 \\
    v^5v_3 & v^4v_1v_2 & v^4v_1v_2 & v^4v_1v_2 & v^3v_1^3 \end{array} \right )
\label{TZsq30}
\eeq
where we use the notations $v_1=1+v$, $v_2=2+v$, $v_3=3+v$ and 
\beqs
s_1 & = & q^3+6q^2v+15qv^2+qv^3+16v^3+3v^4 \cr
s_2 & = & q^2+5qv+qv^2+8v^2+3v^3 \cr
s_3 & = & q+4v+v^2 \cr
s_4 & = & q+3v+v^2 \cr
s_5 & = & q+3v \ .
\eeqs
The eigenvalues of $T_{Z,sq,3,0}$ consist of a linear term $v^2(v+q)(v+1)$ and
roots of a cubic equation. The linear term, denoted as $\lambda_{Z,s3t,10}$ in
\cite{s3a}, has multiplicity two, so the corresponding coefficient is 
$2b^{(0)}=2$. The cubic equation is the same as eqs. (3.48) to (3.51) of
\cite{s3a}. For $d=1$, we have
\beq
T_{Z,sq,3,1} = v \left( \begin{array}{cccccccccc}
s_6 & vs_3 & vs_3 & v_1s_3 & vv_1v_2 & v_1s_4 & v_1s_4 & vv_1v_2 & v_1^3 & vv_1v_2 \\
vs_3 & s_6 & vs_3 & vv_1v_2 & v_1s_4 & vv_1v_2 & v_1s_4 & vv_1v_2 & v_1^3 & v_1s_3 \\
vs_3 & vs_3 & s_6 & vv_1v_2 & v_1s_4 & v_1s_4 & vv_1v_2 & v_1s_3 & v_1^3 & vv_1v_2 \\    
v^3 & 0 & 0 & v^2v_1 & 0 & 0 & 0 & 0 & 0 & 0 \\
v^3v_3 & v^2s_3 & v^2s_3 & v^2v_1v_2 & vv_1s_4 & v^2v_1v_2 & v^2v_1v_2 & v^2v_1v_2 & vv_1^3 & v^2v_1v_2 \\
v^2s_3 & v^3v_3 & v^2s_3 & v^2v_1v_2 & v^2v_1v_2 & vv_1s_4 & v^2v_1v_2 & v^2v_1v_2 & vv_1^3 & v^2v_1v_2 \\
v^2s_3 & v^2s_3 & v^3v_3 & v^2v_1v_2 & v^2v_1v_2 & v^2v_1v_2 & vv_1s_4 & v^2v_1v_2 & vv_1^3 & v^2v_1v_2 \\
0 & 0 & v^3 & 0 & 0 & 0 & 0 & v^2v_1 & 0 & 0 \\
v^4v_3 & v^4v_3 & v^4v_3 & v^3v_1v_2 & v^3v_1v_2 & v^3v_1v_2 & v^3v_1v_2 & v^3v_1v_2 & v^2v_1^3 & v^3v_1v_2 \\
0 & v^3 & 0 & 0 & 0 & 0 & 0 & 0 & 0 & v^2v_1     \end{array} \right )
\label{TZsq31}
\eeq
where
\beq
s_6 = q^2+5qv+8v^2+v^3 \ .
\eeq
The eigenvalues of $T_{Z,sq,3,1}$ consist of the roots of a cubic equation,
which enter with multiplicity two, and the roots of a quartic equation. The
cubic equation is given by eqs. (2.23) to (2.26) in \cite{s3a}, and since it
has multiplicity two, its roots have coefficients $2b^{(1)}=2(q-1)$. The quartic
equation is the same as that given in eqs. (3.43) to (3.47) of \cite{s3a}. For
$d=2$, we have
\beq
T_{Z,sq,3,2} = v^2 \left( \begin{array}{cccccccccccc}
s_5 & 0 & v & 0 & 0 & v & v_1 & 0 & 0 & 0 & v_1 & 0 \\
0 & s_5 & 0 & v & v & 0 & 0 & v_1 & 0 & 0& 0 & v_1 \\
v & 0 & s_5 & 0 & v & 0 & v_1 & 0 & v_1 & 0 & 0 & 0 \\    
0 & v & 0 & s_5 & 0 & v & 0 & v_1 & 0 & v_1 & 0 & 0 \\
0 & v & v & 0 & s_5 & 0 & 0 & 0 & v_1 & 0 & 0 & v_1 \\
v & 0 & 0 & v & 0 & s_5 & 0 & 0 & 0 & v_1 & v_1 & 0 \\
v^2 & 0 & v^2 & 0 & 0 & 0 & vv_1 & 0 & 0 & 0 & 0 & 0 \\
0 & v^2 & 0 & v^2 & 0 & 0 & 0 & vv_1 & 0 & 0 & 0 & 0 \\
0 & 0 & v^2 & 0 & v^2 & 0 & 0 & 0 & vv_1 & 0 & 0 & 0 \\
0 & 0 & 0 & v^2 & 0 & v^2 & 0 & 0 & 0 & vv_1 & 0 & 0 \\
v^2 & 0 & 0 & 0 & 0 & v^2 & 0 & 0 & 0 & 0 & vv_1 & 0 \\
0 & v^2 & 0 & 0 & v^2 & 0 & 0 & 0 & 0 & 0 & 0 & vv_1     \end{array} \right ) \ .
\label{TZsq32}
\eeq
The eigenvalues of $T_{Z,sq,3,2}$ consist of two linear terms and roots of
three quadratic equations; two of the quadratic equations occur with multiplicity
two. Both of the linear terms $v^2(v+q)$ and $v^3(v+1)$ have the coefficient
$b^{(2)}_2=(q-1)(q-2)/2$. One of the quadratic equations is given by eq. (3.39)
of \cite{s3a} with coefficient $q(q-3)/2$. The other two quadratic equations
have multiplicity two, as noted; one of them is given by eq. (2.20) of
\cite{s3a} and its roots have coefficient $2b^{(2)}_2=(q-1)(q-2)$.  The other is given by
eq. (2.21) of \cite{s3a} and its roots have coefficient $2b^{(2)}_1=q(q-3)$.  The matrix
$T_{T,sq,3,3}=v^3$ has been given above and has coefficient $b^{(3)}$.

\subsection{Square-Lattice Strip, $L_{\lowercase{y}}=4$}

For the $L_y=4$ toroidal and Klein bottle strips of the square lattice, the
sizes of matrices are $14 \times 14$, $35 \times 35$, $56 \times 56$, and $48
\times 48$ for levels $0 \le d \le 3$, as indicated in Table
\ref{nztabletor}. The eigenvalues of $T_{Z,sq,4,0}$ are roots of three
quadratic equations and a sixth degree equation. One of the quadratic equations
has multiplicity two. The eigenvalues of $T_{Z,sq,4,1}$ consist of a linear
term $v^3(1+v)(q+v)$, roots of a cubic equation, a sixth-degree equation, an
eighth-degree equation and a ninth-degree equation. The eighth-degree equation
has multiplicity two. The eigenvalues of $T_{Z,sq,4,2}$ consist of a linear
term, roots of two cubic equations, three quartic equations, two sixth-degree
equations and an eighth-degree equation. The linear term is again
$v^3(1+v)(q+v)$ which has multiplicity four with coefficient
$2(q^2-3q+1)$. The eighth-degree equation and one of the sixth-degree equations
have multiplicity two. The eigenvalues of $T_{Z,sq,4,3}$ consist of two linear
terms $v^4(1+v)$ and $v^3(v+q)$, and roots of four quadratic equations and a
quartic equation. The linear terms and one of the quadratic equations have
multiplicity two, and the other equations have multiplicity four. The matrix
$T_{T,sq,4,4}=v^4$ has been given above with coefficient $b^{(4)}$. Some of
these equations are too lengthy to list here; they are available from the
authors. At the special value $v=-1$, the Potts model partition function
reduces to the chromatic polynomial. The non-zero eigenvalues are the same as
$\lambda_{st4,j}$ for $1 \le j \le 33$ from eqs. (2.3) to (2.21) in
\cite{tor4}.

\subsection{Triangular-Lattice Strip, $L_{\lowercase{y}}=2$}

The toroidal strip of the triangular lattice with $L_y=2$ is equivalent to the
ladder graph with next-nearest-neighbor coupling and all transverse edges
doubled. The corresponding Klein bottle strip is the same as the toroidal
strip. For $d=0$, we have
\beq
T_{Z,tri,2,0} = \left( \begin{array}{cc}
    q^2+6qv+qv^2+12v^2+8v^3+2v^4  & (q+4v+2v^2)(1+v)^2 \\
    v^2(2q+12v+13v^2+6v^3+v^4)     & v^2(2+v)^2(1+v)^2 \end{array} \right ) \ .
\label{TZtri20}
\eeq
The eigenvalues of $T_{Z,tri,2,0}$, denoted as $\lambda_{tt2,(1,2)}$, are roots
of the following equation:
\beqs
\xi^2 & - & (v^6+q^2+6v^5+20v^3+15v^4+16v^2+qv^2+6vq)\xi \cr
& & + v^2(1+v)^2 (2v^4+4v^3+4v^3q+11qv^2+q^2v^2+4vq+4q^2v+2q^2) =0 \ .
\eeqs
For $d=1$, we have
\beq
T_{Z,tri,2,1} = v \left( \begin{array}{ccc}
    q+6v+4v^2+v^3 & q+6v+4v^2+v^3 & (2+v)(1+v)^2 \\
    q+6v+4v^2+v^3 & q+6v+4v^2+v^3 & (2+v)(1+v)^2 \\
    v(q+12v+13v^2+6v^3+v^4) & v(q+12v+13v^2+6v^3+v^4) & v(2+v)^2(1+v)^2 \end{array} \right ) \ .
\label{TZtri21}
\eeq
The eigenvalues of $T_{Z,tri,2,1}$ consist of a zero, $\lambda_{tt2,0}=0$, and
roots of the following quadratic equation:
\beq
\xi^2 -v(v^5+6v^4+15v^3+20v^2+16v+2q)\xi + 2v^3(2+v)(1+v)^2(v^2+vq+q)=0
\eeq
with roots $\lambda_{tt2,(3,4)}$. Their coefficients are equal to $b^{(1)}=q-1$. For $d=2$, we have
\beq
T_{Z,tri,2,2} = v^2 \left( \begin{array}{cc}
    1 & 1 \\
    1 & 1 \end{array} \right ) \ .
\label{TZtri22}
\eeq
$T_{Z,tri,2,2}$ has two eigenvalues. One of them is $\lambda_{tt2,5}=2v^2$ with
coefficient $b^{(2)}_1=q(q-3)/2$; while the other eigenvalue is zero with coefficient
$b^{(2)}_2=(q-1)(q-2)/2$. Therefore, the Potts model partition function for the
triangular lattice strip with $L_y=2$ and toroidal boundary condition is given
by
\beq
Z(tri, 2 \times L_x,tor.,q,v) = b^{(0)} \sum_{j=1}^2 (\lambda_{tt2,j})^{L_x} + b^{(1)} \sum_{j=3}^4 (\lambda_{tt2,j})^{L_x} + b_1^{(2)} (\lambda_{tt2,5})^{L_x} \ .
\label{ztritorly2}
\eeq

\subsection{Triangular-Lattice Strip, $L_{\lowercase{y}}=3$}

We illustrate our results for the $L_y=3$ toroidal strip of the triangular
lattice. For $d=0$, we have
\beq
T_{Z,tri,3,0} = \left( \begin{array}{ccccc}
    t_1 & v_1t_2 & v_1t_2 & v_1t_2 & v_1^3t_5 \\
    v^2t_3 & v^2v_1v_2t_4 & v^2v_1v_2t_4 & v^2v_1t_6 & v^2v_1^3v_2^2 \\
    v^2t_3 & v^2v_1t_6 & v^2v_1v_2t_4 & v^2v_1v_2t_4 & v^2v_1^3v_2^2 \\
    v^2t_3 & v^2v_1v_2t_4 & v^2v_1t_6 & v^2v_1v_2t_4 & v^2v_1^3v_2^2 \\
    v^4v_3t_7 & v^3v_1v_2t_8 & v^3v_1v_2t_8 & v^3v_1v_2t_8 & v^3v_1^3v_2^3 \end{array} \right )
\label{TZtri30}
\eeq
where we define 
\beqs
t_1 & = & q^3+9q^2v+33qv^2+4qv^3+50v^3+21v^4+3v^5 \cr
t_2 & = & q^2+8qv+2qv^2+20v^2+14v^3+3v^4 \cr
t_3 & = & q^2+10qv+2qv^2+30v^2+22v^3+7v^4+v^5 \cr
t_4 & = & q+6v+4v^2+v^3 \cr
t_5 & = & q+6v+3v^2 \cr
t_6 & = & 2q+14v+14v^2+6v^3+v^4 \cr
t_7 & = & 3q+18v+15v^2+6v^3+v^4 \cr
t_8 & = & q+12v+13v^2+6v^3+v^4 \ .
\eeqs
The characteristic equation of $T_{Z,tri,3,0}$ yields the following
cubic and quadratic equations: 
\beq
\xi^3 + f_{31}\xi^2 + f_{32}\xi + f_{33} =0
\eeq
where
\beqs
f_{31} & = & -(q^3+v^9+39v^2q+9v^8+36v^7+84v^6+129v^5+137v^4+96v^3+12v^3q+2v^4q \cr & & +9q^2v) \cr\cr 
f_{32} & = & v^2(1+v) (46q^3v+76q^3v^2+246q^2v^5+56v^3q^3+1190v^6q+72q^2v^6+q^3v^6+24v^9q \cr & & +2v^{10}q+3q^4+35v^{10}+3v^{11}+9v^7q^2+534v^7q+147v^8q+483v^3q^2+213v^2q^2 \cr & & +8q^3v^5+28q^3v^4+1272v^4q+2q^4v+414v^3q+794v^5+192v^4+1618v^5q+1091v^7 \cr & & +1243v^6+188v^9+580v^8+455v^4q^2) \cr\cr 
f_{33} & = & -v^5(1+v)^4 (22q^3v+72q^3v^2+105q^2v^5+105v^3q^3+72v^6q+24q^2v^6+6q^4+19v^7q \cr & & +90v^3q^2+30v^2q^2+2v^4q^4+12q^4v^3+24q^4v^2+12q^3v^5+62q^3v^4+60v^4q+19q^4v \cr & & +24v^3q+24v^5+8v^4+90v^5q+22v^7+30v^6+6v^8+148v^4q^2) 
\eeqs
with roots $\lambda_{tt3,j}$ for $1 \le j \le 3$ and
\beq
\xi^2 -v^3(q-2)(1+v)\xi + v^6(q-2)^2(1+v)^2 =0
\eeq
with roots $\lambda_{tt3,j}$ for $j=4, 5$. For $d=1$, we have
\beqs
\lefteqn{T_{Z,tri,3,1} = v \times} \cr\cr
& & \left( \begin{array}{cccccccccc}
t_9 & t_9 & vt_{10} & v_1t_{11} & v_1t_4 & v_1t_4 & v_1v_2s_3 & vv_1v_2^2 & v_1^3v_2 & v_1t_{11} \\
vt_{10} & t_9 & t_9 & vv_1v_2^2 & v_1v_2s_3 & v_1t_4 & v_1t_4 & v_1t_{11} & v_1^3v_2 & v_1t_{11} \\
t_9 & vt_{10} & t_9 & v_1t_{11} & v_1t_4 & v_1v_2s_3 & v_1t_4 & v_1t_{11} & v_1^3v_2 & vv_1v_2^2 \\    
v^2t_{12} & v^2t_{12} & 0 & v^2v_1v_2 & 0 & 0 & v^2v_1v_2 & 0 & 0 & v^2v_1v_2 \\
v^2t_{13} & v^2t_{13} & vt_{14} & v^2v_1v_2t_{15} & vv_1v_2t_4 & vv_1v_2t_4 & vv_1t_8 & vv_1t_{16} & vv_1^3v_2^2 & v^2v_1v_2t_{15} \\
vt_{14} & v^2t_{13} & v^2t_{13} & vv_1t_{16} & vv_1t_8 & vv_1v_2t_4 & vv_1v_2t_4 & v^2v_1v_2t_{15} & vv_1^3v_2^2 & v^2v_1v_2t_{15} \\
v^2t_{13} & vt_{14} & v^2t_{13} & v^2v_1v_2t_{15} & vv_1v_2t_4 & vv_1t_8 & vv_1v_2t_4 & v^2v_1v_2t_{15} & vv_1^3v_2^2 & vv_1t_{16} \\
v^2t_{12} & 0 & v^2t_{12} & v^2v_1v_2 & 0 & v^2v_1v_2 & 0 & v^2v_1v_2 & 0 & 0 \\
v^3v_3t_{17} & v^3v_3t_{17} & v^3v_3t_{17} & v^3v_1v_2v_3t_{18} & v^2v_1v_2t_8 & v^2v_1v_2t_8 & v^2v_1v_2t_8 & v^3v_1v_2v_3t_{18} & v^2v_1^3v_2^3 & v^3v_1v_2v_3t_{18} \\
0 & v^2t_{12} & v^2t_{12} & 0 & v^2v_1v_2 & 0 & 0 & v^2v_1v_2 & 0 & v^2v_1v_2     \end{array} \right ) \cr\cr
& &
\label{TZtri31}
\eeqs
where
\beqs
t_9 & = & q^2+8qv+qv^2+20v^2+8v^3+v^4 \cr
t_{10}& = & 2q+10v+5v^2+v^3 \cr
t_{11}& = & q+8v+5v^2+v^3 \cr
t_{12}& = & q+6v+2v^2 \cr
t_{13}& = & 4q+qv+24v+20v^2+7v^3+v^4 \cr
t_{14}& = & q^2+9qv+2qv^2+30v^2+22v^3+7v^4+v^5 \cr
t_{15}& = & 5+4v+v^2 \cr
t_{16}& = & q+14v+14v^2+6v^3+v^4 \cr
t_{17}& = & 2q+18v+15v^2+6v^3+v^4 \cr
t_{18}& = & 4+3v+v^2 \ .
\eeqs
The characteristic equation of $T_{Z,tri,3,1}$ yields the following quartic
equation and sixth-degree equation:
\beq
\xi^4 + f_{41}\xi^3 + f_{42}\xi^2 + f_{43}\xi + f_{44} =0
\eeq
where
\beqs
f_{41} & = & -v(84v^5+36v^6+9v^7+2v^3q+9v^2q+2q^2+v^8+140v^3+130v^4+98v^2+23vq) \cr\cr
f_{42} & = & v^3(1+v)(4v^{10}+46v^9+2v^9q+242v^8+22v^8q+118v^7q+736v^7+383v^6q+2q^2v^6 \cr & & +1388v^6+797v^5q+16q^2v^5+1638v^5+1071v^4q+1164v^4+56v^4q^2+920v^3q+384v^3 \cr & & +109v^3q^2+396v^2q+141v^2q^2+94q^2v+3q^3v+6q^3) \cr\cr
f_{43} & = & -v^6(v+2)(1+v)^2(33q^3v+48q^3v^2+111q^2v^5+42v^3q^3+834v^6q+19q^2v^6+4v^9q \cr & & +3v^{10}+262v^7q+46v^8q+348v^3q^2+277v^2q^2+3q^3v^5+18q^3v^4+100q^2v+360v^2q \cr & & +1836v^4q+1248v^3q+1277v^5+812v^4+12q^3+1585v^5q+196v^3+602v^7+1127v^6 \cr & & +35v^9+194v^8+268v^4q^2) \cr\cr
f_{44} & = & v^9(v+2)^2(1+v)^5(6v^6+18v^5q+10v^5+34v^4q+18v^4q^2+10v^4+43v^3q^2+4v^3 \cr & & +6v^3q^3+25v^3q+10v^2q+18q^3v^2+34v^2q^2+18q^3v+10q^2v+6q^3)
\eeqs
with roots $\lambda_{tt3,j}$ for $6 \le j \le 9$ and
\beq
\xi^6 + f_{61}\xi^5 + f_{62}\xi^4 + f_{63}\xi^3 + f_{64}\xi^2 + f_{65}\xi + f_{66} =0
\eeq
where
\beqs
f_{61} & = & -v(7v^3+2v^4+12v^2+q^2+v^3q+7vq+3v^2q) \cr\cr
f_{62} & = & v^2(4v^8+168v^4+208v^5+117v^6+32v^7+3v^7q+q^4+q^2v^6+14q^3v+2v^3q^3+6q^3v^2 \cr & & +73v^2q^2+21v^6q+26v^4q^2+168v^4q+172v^3q+55v^3q^2+6q^2v^5+84v^5q) \cr\cr
f_{63} & = & -v^6(1+v)(2v^7q-8v^7+11v^6q+q^2v^6-60v^6+24v^5q+7q^2v^5-200v^5-10v^4q \cr & & -336v^4+27v^4q^2+2v^3q^3+47v^3q^2-116v^3q-272v^3-200v^2q+7q^3v^2+32v^2q^2 \cr & & +13q^3v-48q^2v+q^4-4q^3) \cr\cr
f_{64} & = & v^8(1+v)^2(22q^3v+9q^3v^2+38q^2v^5+17v^3q^3+v^8q^2-212v^6q+28q^2v^6+2q^4 \cr & & +8v^7q^2-64v^7q-8v^8q-14v^3q^2+104v^2q^2+q^4v^2+2q^3v^5+9q^3v^4-4v^4q+q^4v \cr & & +344v^3q+872v^5+672v^4-288v^5q+136v^7+492v^6+16v^8+3v^4q^2) \cr\cr
f_{65} & = & v^{12}(q-4)(v+2)(1+v)^3(v^3+4v^2+6v+q)(q+2v)^2 \cr\cr
f_{66} & = & v^{14}(v+2)^2(1+v)^4(q+2v)^4
\eeqs
with roots $\lambda_{tt3,j}$ for $10 \le j \le 15$. Their coefficients are equal to $b^{(1)}=q-1$. For $d=2$, we have
\beq
T_{Z,tri,3,2} = v^2 \left( \begin{array}{cccccccccccc}
t_{19} & v & t_{12} & 0 & t_{19} & v & v_1v_2 & 0 & v_1 & v_1 & v_1v_2 & 0 \\
v & t_{19} & 0 & t_{12} & v & t_{19} & 0 & v_1v_2 & v_1 & v_1 & 0 & v_1v_2 \\
v & t_{19} & t_{19} & v & t_{12} & 0 & v_1 & v_1 & 0 & v_1v_2 & v_1v_2 & 0 \\  
t_{19} & v & v & t_{19} & 0 & t_{12} & v_1 & v_1 & v_1v_2 & 0 & 0 & v_1v_2 \\
0 & t_{12} & v & t_{19} & t_{19} & v & 0 & v_1v_2 & 0 & v_1v_2 & v_1 & v_1 \\
t_{12} & 0 & t_{19} & v & v & t_{19} & v_1v_2 & 0 & v_1v_2 & 0 & v_1 & v_1 \\
v^2v_3 & v^2v_3 & vt_{12} & 0 & vt_{12} & 0 & vv_1v_2 & 0 & 0 & vv_1v_2 & vv_1v_2 & 0 \\
v^2v_3 & v^2v_3 & 0 & vt_{12} & 0 & vt_{12} & 0 & vv_1v_2 & vv_1v_2 & 0 & 0 & vv_1v_2 \\
vt_{12} & 0 & vt_{12} & 0 & v^2v_3 & v^2v_3 & vv_1v_2 & 0 & vv_1v_2 & 0 & vv_1v_2 & 0 \\
0 & vt_{12} & 0 & vt_{12} & v^2v_3 & v^2v_3 & 0 & vv_1v_2 & 0 & vv_1v_2 & 0 & vv_1v_2 \\
0 & vt_{12} & v^2v_3 & v^2v_3 & vt_{12} & 0 & 0 & vv_1v_2 & 0 & vv_1v_2 & vv_1v_2 & 0 \\
vt_{12} & 0 & v^2v_3 & v^2v_3 & 0 & vt_{12} & vv_1v_2 & 0 & vv_1v_2 & 0 & 0 & vv_1v_2     \end{array} \right )
\label{TZtri32}
\eeq
where
\beq
t_{19} = q+5v+v^2 \ .
\eeq
The characteristic equation $T_{Z,tri,3,2}$ yields two linear equations, a
quadratic equation and two quartic equations. The eigenvalues which are the
solutions of the two linear equations, 
$\lambda_{tt3,16}=v^3(v+1)(v+2)$ and $\lambda_{tt3,17}=v^2(2v+q)$, have the
coefficient $b^{(2)}_2=(q-1)(q-2)/2$. One of the quartic equations is given by
\beqs
& & \xi^4 -v^2(9v^2+2v^3+14v+2q)\xi^3 \cr & & + v^4(36v^5+135v^4+246v^3+192v^2+4v^6+7v^3q+33v^2q+54vq+4q^2)\xi^2 \cr & & -v^7(v+2)(1+v)(9v^2+2v^3+14v+2q)(q+2v)\xi + v^{10}(v+2)^2(1+v)^2(q+2v)^2 =0 \cr & &
\eeqs
with roots $\lambda_{tt3,j}$ for $18 \le j \le 21$ that also have $b^{(2)}_2$
as their coefficient. The roots of the following two equations have coefficient
$b^{(2)}_1=q(q-3)/2$:
\beq
\xi^2 -v^2(3q+24v+13v^2+3v^3)\xi + v^5(1+v)(6v^2+5vq+6q) =0
\eeq
with roots $\lambda_{tt3,(22,23)}$ and
\beq
\xi^4 + v^4\xi^3 + v^6(q+vq+v^2)\xi^2 -v^{10}q(1+v)\xi + v^{12}q^2(1+v)^2 =0
\eeq
with roots $\lambda_{tt3,j}$ for $24 \le j \le 27$. For $d=2$, we have
\beq
T_{Z,tri,3,3} = v^3 \left( \begin{array}{cccccc}
    1 & 0 & 0 & 1 & 0 & 0 \\
    0 & 1 & 1 & 0 & 0 & 0 \\
    0 & 0 & 1 & 0 & 0 & 1 \\
    0 & 0 & 0 & 1 & 1 & 0 \\
    1 & 0 & 0 & 0 & 1 & 0 \\
    0 & 1 & 0 & 0 & 0 & 1 \end{array} \right ) \ .
\label{TZtri33}
\eeq
All the eigenvalues of $T_{Z,tri,3,3}$ have multiplicity two. These eigenvalues
are $\lambda_{tt3,28}=2v^3$ with coefficient $b^{(3)}_1=(q-1)(q^2-5q+3)/3$ and
the roots of $\xi^2-v^3\xi+v^6=0$, denoted as $\lambda_{tt3,(29,30)}$, with
coefficient $2b^{(3)}_2=q(q-2)(q-4)/3$. Therefore, the Potts model partition
function for the triangular lattice strip with $L_y=3$ and toroidal boundary
condition is given by
\beqs
Z(tri, 3 \times L_x,tor.,q,v) & = & b^{(0)} \sum_{j=1}^5 (\lambda_{tt3,j})^{L_x} + b^{(1)} \sum_{j=6}^{15} (\lambda_{tt3,j})^{L_x} + b_2^{(2)} \sum_{j=16}^{21} (\lambda_{tt3,j})^{L_x} \cr\cr & & + b_1^{(2)} \sum_{j=22}^{27} (\lambda_{tt3,j})^{L_x} + b^{(3)}_1 (\lambda_{tt3,28})^{L_x} + 2b^{(3)}_2 \sum_{j=29}^{30} (\lambda_{tt3,j})^{L_x} \ . \quad
\label{ztritorly3}
\eeqs
For the corresponding Klein bottle strip, the coefficients for
$\lambda_{tt3,j}$ with $10 \le j \le 15$, $18 \le j \le 21$, $24 \le j \le 27$
and $j=4,5,29,30$ become zero. The coefficient for $\lambda_{tt3,17}$ changes
sign and the coefficient for $\lambda_{tt3,28}$ becomes
$-b^{(1)}=1-q$. Consequently, the Potts model partition function for the
triangular lattice strip with $L_y=3$ and Klein bottle boundary condition is
given by
\beqs
Z(tri, 3 \times L_x,Kb.,q,v) & = & b^{(0)} \sum_{j=1}^3 (\lambda_{tt3,j})^{L_x} + b^{(1)} \left ( \sum_{j=6}^9 (\lambda_{tt3,j})^{L_x} - (\lambda_{tt3,28})^{L_x} \right ) \cr\cr & & + b_2^{(2)} \left ( (\lambda_{tt3,16})^{L_x} - (\lambda_{tt3,17})^{L_x} \right ) + b_1^{(2)} \sum_{j=22}^{23} (\lambda_{tt3,j})^{L_x} \ .
\label{ztrikbly3}
\eeqs
At the special value $v=-1$, the Potts model partition function reduces to the
chromatic polynomial. The non-zero eigenvalues are the same as $\lambda_{tt,j}$
for $1 \le j \le 11$ from eqs. (4.2) to (4.9) in \cite{t}.

\subsection{Honeycomb-Lattice Strip, $L_{\lowercase{y}}=4$}

For the $L_y=4$ toroidal and Klein bottle strips of the honeycomb lattice, the
linear sizes of the matrices are 14, 35, 56, 48 for levels $0 \le d \le 3$, as
indicated in Table \ref{nztabletor}. The eigenvalues of $T_{Z,hc,4,0}$ are
roots of two quadratic equations and an eighth-degree equation. One of the 
quadratic equations has multiplicity two. The eigenvalues of
$T_{Z,hc,4,1}$ are roots of a seventh-degree equation, an eighth-degree
equation and a twelfth-degree equation. The eighth-degree equation has 
multiplicity two. The eigenvalues of $T_{Z,hc,4,2}$ are roots of a quartic
equation, a fifth-degree equation, a sixth-degree equation, two eighth-degree
equations and a eleventh-degree equation. The sixth-degree equation and one of
the eighth-degree equations have multiplicity two. The eigenvalues of
$T_{Z,hc,4,3}$ consist of a linear term $v^6(v+q)^2$, roots of a quadratic
equation, a cubic equation and two quartic equations. The linear terms and the
cubic equation have multiplicity two, and the other equations have multiplicity
four. The matrix $T_{T,hc,4,4}=v^8$ has been given above with coefficient
$b^{(4)}$. Some of these equations are too lengthy to list here. They are
available from the authors.

At the special value $v=-1$, the Potts model partition function reduces to the
chromatic polynomial. The linear sizes of the matrices reduce to 6, 18, 34, 36
for levels $0 \le d \le 3$, as indicated in Table \ref{nptabletorhc}. The
eigenvalues of $T_{P,hc,4,0}$ consist of two linear terms and roots of a
quartic equation. They are given by
\beq
\lambda_{ht4,0} = 0 
\label{lzero}
\eeq
\beq
\lambda_{ht4,1} = (q-1)^2(q-3)^2 
\label{l0}
\eeq
\beqs
& & \xi^4 - (q^8-12q^7+66q^6-220q^5+496q^4-796q^3+922q^2-734q+314)\xi^3 \cr & & + (q^{12}-20q^{11}+188q^{10}-1092q^9+4344q^8-12428q^7+26192q^6-41022q^5+47597q^4 \cr & & -40318q^3+24247q^2-9782q+2169)\xi^2 \cr & & - (4q^{12}-80q^{11}+736q^{10}-4124q^9+15705q^8-42922q^7+86535q^6-129964q^5 \cr & & +144575q^4-116374q^3+64525q^2-22280q+3680)\xi + 4(q-1)^4(q-2)^4=0
\label{eq0}
\eeqs
with roots $\lambda_{ht4,j}$ for $2 \le j \le 5$. $\lambda_{ht4,0}=0$ appears
as one of the eigenvalues of $T_{P,hc,4,1}$ with multiplicity two. The
remaining eigenvalues of $T_{P,hc,4,1}$ are roots of a cubic equation, a
quartic equation and a sixth-degree equation. The cubic equation has
multiplicity two. These equations are given by
\beqs
& & \xi^3 - (q^6-10q^5+44q^4-110q^3+167q^2-150q+68)\xi^2 \cr & & + (q^8-14q^7+90q^6-342q^5+832q^4-1322q^3+1345q^2-812q+230)\xi \cr & & - (q-1)^2(2q^6-22q^5+102q^4-256q^3+373q^2-308q+115) =0
\label{eq11}
\eeqs
with roots $\lambda_{ht4,j}$ for $j=6,7,8$.
\beqs
& & \xi^4 - (q^6-8q^5+29q^4-64q^3+95q^2-86q+37)\xi^3 \cr & & + (q-1)^2(q^8-14q^7+86q^6-302q^5+672q^4-1000q^3+1030q^2-724q+280)\xi^2 \cr & & - (q-1)^4(q^8-16q^7+112q^6-444q^5+1084q^4-1664q^3+1581q^2-886q+253)\xi \cr & & + (q-3)^2(q-1)^6=0
\label{eq12}
\eeqs
with roots $\lambda_{ht4,j}$ for $9 \le j \le 12$.
\beqs
&  & \xi^6 - (q^6-12q^5+65q^4-204q^3+407q^2-510q+321)\xi^5 + (q^{10}-20q^9+187q^8 \cr & & -1064q^7+4046q^6-10670q^5+19694q^4-25280q^3+22278q^2-13120q+4368)\xi^4 \cr & & - (q^{12}-24q^{11}+274q^{10}-1948q^9+9541q^8-33732q^7+87865q^6-169306q^5+239311q^4 \cr & & -243300q^3+172063q^2-79202q+19177)\xi^3 + (4q^{12}-92q^{11}+992q^{10}-6588q^9 \cr & & +29885q^8-97284q^7+232764q^6-412788q^5+540290q^4-512128q^3+336720q^2 \cr & & -139512q+27993)\xi^2 -4(q-1)^2(q^{10}-20q^9+182q^8-984q^7+3483q^6-8402q^5 \cr & & +13956q^4-15756q^3+11602q^2-5078q+1032)\xi + 16(q-2)^2(q-1)^6=0 
\label{eq13}
\eeqs
with roots $\lambda_{ht4,j}$ for $13 \le j \le 18$. $\lambda_{ht4,0}=0$ is
again one of the eigenvalues of $T_{P,hc,4,2}$. The remaining eigenvalues of
$T_{P,hc,4,2}$ are roots of a quadratic equation, two cubic equations, two
fifth-degree equations and a seventh-degree equation. They are given by
\beq
\lambda_{ht4,(19,20)} = \frac12 \left [ q^2-4q+7 \pm (q-3) (q^2-2q+5)^{1/2} \right ] 
\label{eq21}
\eeq
\beqs
& & \xi^5 - (2q^4-14q^3+41q^2-60q+41)\xi^4 + (q^8-14q^7+91q^6-358q^5+932q^4-1636q^3 \cr & & +1875q^2-1270q+396)\xi^3 - (q^{10}-18q^9+148q^8-726q^7+2339q^6-5148q^5+7824q^4 \cr & & -8136q^3+5615q^2-2372q+477)\xi^2 + (q-1)^2(q^8-14q^7+85q^6-292q^5+626q^4 \cr & & -874q^3+800q^2-448q+125)\xi -4(q-1)^4=0
\label{eq24}
\eeqs
with roots $\lambda_{ht4,j}$ for $21 \le j \le 25$.
\beqs
& & \xi^5 - (2q^4-15q^3+51q^2-90q+73)\xi^4 + (q^8-15q^7+104q^6-429q^5+1152q^4-2064q^3 \cr & & +2415q^2-1694q+570)\xi^3 - (q^{10}-19q^9+167q^8-881q^7+3070q^6-7372q^5+12396q^4 \cr & & -14520q^3+11459q^2-5550q+1265)\xi^2 + (q-1)^2(q^8-15q^7+104q^6-428q^5+1144q^4 \cr & & -2039q^3+2392q^2-1706q+575)\xi -2q(q-1)^6=0
\label{eq25}
\eeqs
with roots $\lambda_{ht4,j}$ for $26 \le j \le 30$.
\beqs
& & \xi^3 - (q^4-7q^3+21q^2-32q+24)\xi^2 + (q^6-11q^5+53q^4-139q^3+206q^2-164q+58)\xi \cr & & -(q-1)^2(q^4-7q^3+18q^2-24q+15) =0
\label{eq23}
\eeqs
with roots $\lambda_{ht4,j}$ for $j=31,32,33$.
\beqs
& & \xi^3 - (q^4-8q^3+27q^2-40q+24)\xi^2 + (q-1)^2(q^4-10q^3+44q^2-90q+72)\xi \cr & & -(q-3)^2(q-1)^4 =0
\label{eq22}
\eeqs
with roots $\lambda_{ht4,j}$ for $j=34,35,36$.
\beqs
& & \xi^7 - (3q^4-22q^3+75q^2-140q+126)\xi^6 + (3q^8-44q^7+304q^6-1282q^5+3601q^4 \cr & & -6890q^3+8830q^2-7038q+2781)\xi^5 - (q^{12}-22q^{11}+230q^{10}-1504q^9+6839q^8 \cr & & -22756q^7+56785q^6-107160q^5+152314q^4-160280q^3+120302q^2-59250q+15045)\xi^4 \cr & & + (q^{14}-26q^{13}+318q^{12}-2414q^{11}+12685q^{10}-48806q^9+141948q^8-317702q^7 \cr & & +551665q^6-743122q^5+769969q^4-601782q^3+341290q^2-128886q+25246)\xi^3 \cr & & - (q^{14}-24q^{13}+280q^{12}-2088q^{11}+11028q^{10}-43286q^9+129310q^8-297078q^7 \cr & & +525432q^6-710236q^5+721717q^4-535598q^3+276095q^2-89634q+14145)\xi^2 \cr & & + 4(q-1)^2(q^9-6q^8-14q^7+249q^6-1080q^5+2522q^4-3564q^3+3069q^2-1499q \cr & & +326)\xi -4(q-2)^2(q-1)^6=0
\label{eq26}
\eeqs
with roots $\lambda_{ht4,j}$ for $37 \le j \le 43$. The second fifth- degree
equation and the first cubic equation given above have multiplicity two.
$\lambda_{ht4,0}=0$ is again one of the eigenvalues of $T_{P,hc,4,3}$ with
multiplicity two. The remaining eigenvalues of $T_{P,hc,4,3}$ are two linear
terms, roots of a quadratic equation and two cubic equations. They are given by
\beq
\lambda_{ht4,44} = q^2-4q+5 
\label{l32}
\eeq
\beq
\lambda_{ht4,45} = (q-1)^2 
\label{l31}
\eeq
\beq
\lambda_{ht4,(46,47)} = \frac12 \left [ q^2-6q+13 \pm (q-3) (q^2-6q+17)^{1/2} \right ] 
\label{eq31}
\eeq
\beq
\xi^3 - (2q^2-8q+11)\xi^2 + (q^4-8q^3+26q^2-38q+23)\xi -(q-1)^2 =0
\label{eq32}
\eeq
with roots $\lambda_{ht4,j}$ for $j=48,49,50$.
\beq
\xi^3 - (2q^2-8q+13)\xi^2 + (q^4-8q^3+26q^2-34q+19)\xi -3(q-1)^2 =0
\label{eq33}
\eeq
with roots $\lambda_{ht4,j}$ for $j=51,52,53$. $\lambda_{ht4,j}$ for
$j=45,46,47$ have multiplicity two, and the other eigenvalues of $T_{P,hc,4,3}$
have multiplicity four. Finally, we have $\lambda_{ht4,54}=1$ as the eigenvalue
of $T_{P,hc,4,4}$. The corresponding coefficients are
\beqs
b_{ht4,j} & = & 1 \qquad \mbox{for} \ 1 \le j \le 5 \cr\cr
b_{ht4,j} & = & 2(q-1) \qquad \mbox{for} \ 6 \le j \le 8 \cr\cr 
b_{ht4,j} & = & q-1 \qquad \mbox{for} \ 9 \le j \le 18 \cr\cr 
b_{ht4,j} & = & \frac12(q-1)(q-2) \qquad \mbox{for} \ 19 \le j \le 25 \cr\cr 
b_{ht4,j} & = & (q-1)(q-2) \qquad \mbox{for} \ 26 \le j \le 30 \cr\cr 
b_{ht4,j} & = & q(q-3) \qquad \mbox{for} \ 31 \le j \le 33 \cr\cr 
b_{ht4,j} & = & \frac12q(q-3) \qquad \mbox{for} \ 34 \le j \le 43 \cr\cr 
b_{ht4,44} & = & \frac23(q-1)(q^2-5q+3) \cr\cr
b_{ht4,j} & = & \frac13(q-1)(q^2-5q+3) \qquad \mbox{for} \ 45 \le j \le 47 \cr\cr 
b_{ht4,j} & = & \frac23q(q-2)(q-4) \qquad \mbox{for} \ 48 \le j \le 53 \cr\cr 
b_{ht4,54} & = & q^4-8q^3+20q^2-15q+1 \ .
\eeqs
Although $\lambda_{ht4,0}=0$ does not contribute the the chromatic polynomial,
we can still calculate its coefficient for the toroidal strip, and we get
$(q-1)(2q^2-7q+12)/6$. The sum of all the coefficients is equal to
$q^2(q-1)^2$, as dictated by the $L_y=4$ special case of the general result
(\ref{cpsumtorhc}). Hence, the chromatic polynomial for the honeycomb
lattice strip with $L_y=4$ and toroidal boundary condition is given by
\beqs 
P(hc, 4 \times L_x,tor.,q) & = & b^{(0)} \sum_{j=1}^5 (\lambda_{ht4,j})^m
+ b^{(1)} \left ( 2 \sum_{j=6}^{8} (\lambda_{ht4,j})^m + \sum_{j=9}^{18}
(\lambda_{ht4,j})^m \right ) \cr\cr & & + b_2^{(2)} \left ( \sum_{j=19}^{25}
(\lambda_{ht4,j})^m + 2 \sum_{j=26}^{30} (\lambda_{ht4,j})^m \right ) \cr\cr &
& + b_1^{(2)} \left ( 2 \sum_{j=31}^{33} (\lambda_{ht4,j})^m + \sum_{j=34}^{43}
(\lambda_{ht4,j})^m \right ) \cr\cr & & + b^{(3)}_1 \left ( 2
(\lambda_{ht4,44})^m + \sum_{j=45}^{47} (\lambda_{ht4,j})^m \right ) \cr\cr & &
+ 4b^{(3)}_2 \sum_{j=48}^{53} (\lambda_{ht4,j})^m + b^{(4)}
(\lambda_{ht4,54})^m
\label{zhctorly4}
\eeqs
where $m=L_x/2$ as given in eq. (\ref{lxm}). For the corresponding Klein bottle
strip, $\lambda_{ht4,j}$ for $6 \le j \le 8$, $26 \le j \le 33$, $48 \le j \le
53$ and $j=44$ do not contribute. The coefficient for $\lambda_{ht4,j}$ with $9
\le j \le 12$, $21 \le j \le 25$, $34 \le j \le 36$ and $j=1$ changes sign. The
coefficient for $\lambda_{ht4,45}$ becomes $b^{(1)}=q-1$, the coefficients for
$\lambda_{ht4,(46,47)}$ become $-b^{(1)}=1-q$ and the coefficients for
$\lambda_{ht4,54}$ become one. Therefore, the chromatic polynomial for
the honeycomb lattice strip with $L_y=4$ and Klein bottle boundary condition is
given by
\beqs
P(hc, 4 \times L_x,Kb.,q) & = & b^{(0)} \left ( - (\lambda_{ht4,1})^m + \sum_{j=2}^5 (\lambda_{ht4,j})^m + (\lambda_{ht4,54})^m \right ) \cr\cr & & + b^{(1)} \left ( - \sum_{j=9}^{12} (\lambda_{ht4,j})^m + \sum_{j=13}^{18} (\lambda_{ht4,j})^m + (\lambda_{ht4,45})^m - \sum_{j=46}^{47} (\lambda_{ht4,j})^m \right ) \cr\cr & & + b_2^{(2)} \left ( \sum_{j=19}^{20} (\lambda_{ht4,j})^m - \sum_{j=21}^{25} (\lambda_{ht4,j})^m \right ) \cr\cr & & + b_1^{(2)} \left (- \sum_{j=34}^{36} (\lambda_{ht4,j})^m + \sum_{j=37}^{43} (\lambda_{ht4,j})^m \right ) \ .
\label{zhckbly4}
\eeqs

\begin{figure}[hbtp]
\centering
\leavevmode
\epsfxsize=3.0in
\begin{center}
\leavevmode
\epsffile{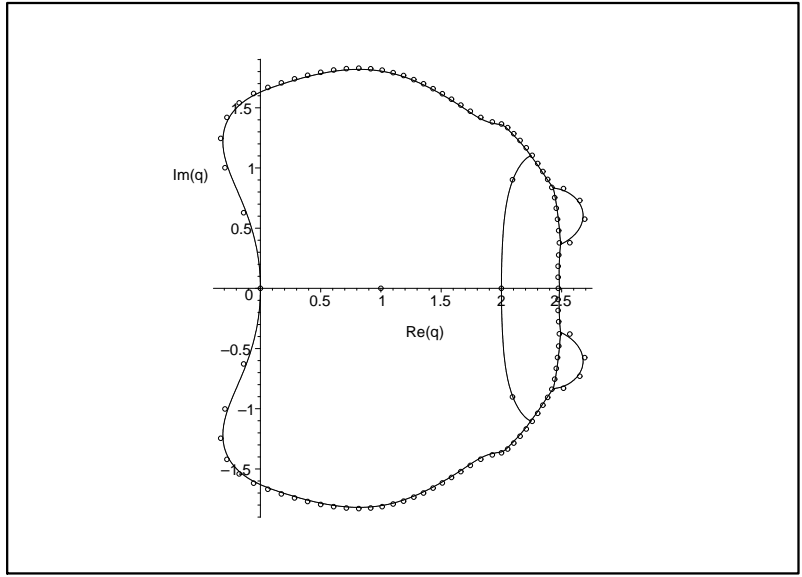}
\end{center}
\vspace*{-2cm}
\caption{\footnotesize{Singular locus ${\cal B}$ for the $L_x \to \infty$ limit
of the strip of the honeycomb lattice with $L_y=4$ and toroidal boundary
conditions. For comparison, chromatic zeros are shown for $L_x=13$ (i.e., 
$n=104$).}}
\label{hcpxpy4}
\end{figure}

The singular locus ${\cal B}$ for the $L_x \to \infty$ limit of the strip of
the honeycomb lattice with $L_y=4$ and toroidal boundary conditions is shown in
Fig. \ref{hcpxpy4}.  For comparison, chromatic zeros are calculated and shown 
for length $L_x=13$ (i.e., $n=104$ vertices). 
The locus ${\cal B}$ crosses the real axis at the points $q=0$, $q=2$, and at 
the maximal point $q=q_c$, where
\beq
q_c \simeq 2.480749  \quad {\rm for} \ \ 
\{G\}=(hc, 4 \times \infty,PBC_y,PBC_x) \ .
\label{qchc}
\eeq
We note that this is rather close to the value for the infinite 2D honeycomb
lattice, namely $q_c(hc)=(3+\sqrt{5} \ )/2 \simeq 2.61803$. As is evident from
Fig. \ref{hcpxpy4}, the locus ${\cal B}$ separates the $q$ plane into different
regions including the following: (i) region $R_1$, containing the semi-infinite
intervals $q > q_c$ and $q < 0$ on the real axis and extending outward to
infinite $|q|$; (ii) $R_2$ containing the interval $2 < q < q_c$; (iii) $R_3$
containing the real interval $0 < q < 2$; and (iv) the complex-conjugate pair
$R_4,R_4^*$ centered approximately at $q=2.55 \pm 0.6i$.  The (nonzero)
densities of chromatic zeros have the smallest values on the curve separating
regions $R_1$ and $R_3$ in the vicinity of the point $q=0$ and on the curve
separating regions $R_2$ and $R_3$ in the vicinity of the point $q=2$.

It is instructive to compare this locus ${\cal B}$ in Fig. \ref{hcpxpy4} with
the corresponding loci that we obtained previously for strips of the honeycomb
lattice with periodic longitudinal boundary conditions and free or periodic
transverse boundary conditions.  We recall that for a given infinite-length
lattice strip with specified transverse boundary conditions, the locus ${\cal
B}$ is the same for periodic longitudinal and twisted periodic longitudinal
boundary conditions (i.e., cyclic versus M\"obius if the transverse B.C. are
free, and toroidal versus Klein bottle if the transverse B.C. are periodic).
For the cyclic strip of the honeycomb lattice, ${\cal B}$ was given as (i)
Fig. 1 of Ref. \cite{pg} for width $L_y=2$, and (ii) Figs. 7, 8, and 9 of
\cite{pt} for $L_y=3,4,5$ (see also the zeros in Fig. 17 of Ref. \cite{hca} for
$L_y=3$). We recall that $L_y$ must be even for strips of the honeycomb lattice
with toroidal boundary conditions.  Comparing our results for ${\cal B}$ for
the width $L_y=4$ toroidal strip of the honeycomb lattice in
Fig. \ref{hcpxpy4} with our earlier results for the $L_y=4$ cyclic strip of
this lattice in Fig. 8 of \cite{pt}, we see that the locus ${\cal B}$ is
significantly simpler for toroidal, versus cyclic, boundary conditions, with
fewer regions and, in particular, no evidence of the tiny sliver phases that
are present in the latter case. Furthermore, the value of $q_c \simeq 2.481$
that we obtain (eq. (\ref{qchc})) for the width $L_y=4$ infinite-length strip of the honeycomb lattice with toroidal boundary conditions is closer to the value of $q_c(hc) \simeq 2.618$ for the infinite 2D than the value $q_c \simeq 2.155$ for the corresponding cyclic $L_y=4$ strip. This is in agreement with one's general expectations, since toroidal B.C. minimize finite-size effects to a greater extent than cyclic B.C. and hence expedite the approach to the $L_y \to \infty$ limit.

In region $R_1$, the dominant eigenvalue is the root of eq. (\ref{eq0}) with
largest magnitude. Denoting it as $\lambda_{ht4,2}$, we have 
\beq
W = (\lambda_{ht4,2})^{1/8} \ , \quad q \in R_1 \ . 
\label{whcr1}
\eeq
This is the same as $W$ for the corresponding $L_x \to \infty$ limit of the
strip of the honeycomb lattice with the same width $L_y=4$ and cylindrical
$(PBC_y,FBC_x)$ boundary conditions, calculated in \cite{hca}.  This equality
of the $W$ functions for the $L_x \to \infty$ limit of two strips of a given
lattice with the same transverse boundary conditions and different longitudinal
boundary conditions in the more restrictive region $R_1$ defined by the two
boundary conditions is a general result \cite{wcy,bcc}. In \cite{hca} we have
listed values of $W$ for a range of values of $q$ for the $L_x \to \infty$
limit of various strips of the honeycomb lattice, including $(hc, 4 \times
\infty,PBC_y,FBC_x)$.  Since $W$ is independent of $BC_x$ for $q$ in the more
restrictive region $R_1$ defined by $FBC_x$ and $(T)PBC_x$ (which is the $R_1$
defined by $PBC_x$ here), it follows, in particular, that \beq W(4 \times
\infty,PBC_y,(T)PBC_x,q)=W(4 \times \infty,PBC_y,FBC_x,q) \quad {\rm for} \ \ q
\ge 2.48...
\label{whctor4qgeqc}
\eeq

In region $R_2$, the largest root of the seventh degree equation (\ref{eq26})
is dominant; we label this as $\lambda_{ht4,37}$, so that
\beq
|W| = |\lambda_{ht4,37}|^{1/8} \ , \quad q \in R_2
\label{whcr2}
\eeq
(in regions other than $R_1$, only $|W|$ can be determined unambiguously 
\cite{w}). Thus, $q_c$ is the relevant solution of the equation of degeneracy
in magnitude $|\lambda_{ht4,2}|=|\lambda_{ht4,37}|$.  In region $R_3$,
\beq
|W|=|\lambda_{ht4,13}|^{1/8} \ , \quad q \in R_3
\label{whcr3}
\eeq 
where $\lambda_{ht4,13}$ is the largest root of the sixth degree equation
(\ref{eq13}). In regions $R_4,R_4^*$,
\beq
|W|=|\lambda_{ht4,9}|^{1/8} \ , \quad q \in R_4, \ R_4^*
\label{whcr4}
\eeq
where $\lambda_{ht4,9}$ is the largest root of the quartic equation
(\ref{eq12}). We calculate the following specific values: $|W(q=1)|=1.67290..$,
$|W(q=2)|=1.35817..$, and $|W(q=q_c)|=1.26949..$ These values may be compared
with those listed in Ref. \cite{hca}.  As is evident from Fig. \ref{hcpxpy4}, the curve ${\cal B}$ has
support for $Re(q) < 0$. For strips with finite lengths $m \ge 3$, certain
chromatic zeros have support for $Re(q) < 0$.

\section{Summary}

In summary, in this paper we have presented a method for calculating transfer
matrices for the $q$-state Potts model partition functions $Z(G,q,v)$, for
arbitrary $q$ and temperature variable $v$, on strip graphs $G$ of the square,
triangular, and honeycomb lattices with width $L_y$ and arbitrarily great
length $L_x$, having toroidal and Klein bottle boundary conditions.  Using this
method, we have derived a number of general properties of these transfer
matrices.  In particular, we have found some very simple formulas for the
determinant $det(T_{Z,\Lambda,L_y,d})$, and trace $Tr(T_{Z,\Lambda,L_y})$.
A number of explicit exact calculations were given.

\bigskip

Acknowledgments: The research of R.S. was partially
supported by the NSF grant PHY-00-98527.  The research of S.C.C. was partially
supported by the NSC grant NSC-93-2119-M-006-009.

\vfill
\eject

\newpage

\begin{thebibliography}{99}

\bibitem{potts}
R. B. Potts, Proc. Camb. Phil. Soc. {\bf 48} (1952) 106.

\bibitem{wurev}
F. Y. Wu, Rev. Mod. Phys. {\bf 54} (1982) 235.

\bibitem{a}
R. Shrock, Physica A {\bf 283} (2000) 388.

\bibitem{bcc}
R. Shrock, in the {\it Proceedings of the 1999 British Combinatorial 
Conference, BCC99} (July, 1999); Physica A {\bf 281}, 221 (2000).

\bibitem{bbook}
N. L. Biggs, {\it Algebraic Graph Theory} (Cambridge
Univ. Press, Cambridge, 1st ed. 1974, 2nd ed. 1993).

\bibitem{rtrev}
R. C. Read and W. T. Tutte, ``Chromatic Polynomials'',
in {\it Selected Topics in Graph Theory, 3}, (Academic Press, New York, 1988),
p. 15.

\bibitem{w}
R. Shrock and S.-H. Tsai, Phys. Rev. {\bf E55} (1997) 5165.

\bibitem{s3a}
S.-C. Chang and R. Shrock, Physica A {\bf 296} (2001) 234. 

\bibitem{ka3}
S.-C. Chang and R. Shrock, in {\it Proc. of the CRM Workshop on Tutte
Polynomials, 2001}, Advances in Applied. Math. {\bf 32} (2004) 44.

\bibitem{bds}
N. L. Biggs, R. M. Damerell, and D. A. Sands, J. Combin. Theory
B {\bf 12} (1972) 123.

\bibitem{bm}
N. L. Biggs and G. H. Meredith, J. Combin. Theory B {\bf 20} (1976) 5.

\bibitem{baxter86}
R. J. Baxter, J. Phys. A {\bf 19} (1986) 2821.

\bibitem{baxter87}
R. J. Baxter, J. Phys. A {\bf 20} (1987) 5241.

\bibitem{matmeth}
N. L. Biggs, J. Combin. Theory B {\bf 82} (2001) 19.

\bibitem{matmeth2}
N. L. Biggs, Bull. London Math. Soc. {\bf 34} (2002) 129.

\bibitem{matmeth3}
N. L. Biggs, London School of Economics Centre for Discrete and Applicable
Mathematics, report LSE-CDAM-2000-04.

\bibitem{sqtran}
J. Salas and A. Sokal, J. Stat. Phys. {\bf 104} (2001) 609.

\bibitem{cyltran}
J.Jacobsen and J. Salas, J. Stat. Phys. {\bf 104} (2001) 701.

\bibitem{tritran}
J. Jacobsen, J. Salas, and A.D. Sokal, J. Stat. Phys. {\bf 112} (2003) 921.

\bibitem{ts}
S.-C. Chang, J. Salas, and R. Shrock, J. Stat. Phys. {\bf 107} (2002) 1207.

\bibitem{tt}
S.-C. Chang, J. Jacobsen, J. Salas, and R. Shrock,
J. Stat. Phys. {\bf 114} (2004) 763.

\bibitem{pt}
S.-C. Chang and R. Shrock, Physica A {\bf 346} (2005) 400.

\bibitem{zt}
S.-C. Chang and R. Shrock, Physica A {\bf 347} (2005) 314.

\bibitem{salas04}
J. Jacobsen and J. Salas, cond-mat/0407444.

\bibitem{pg}
R. Shrock and S.-H. Tsai, J. Phys. A Lett. {\bf 32}, L195-L200 (1999)

\bibitem{pm}
R. Shrock, Phys. Lett. A {\bf 261} (1999) 57.

\bibitem{uspensky}
J. V. Uspensky, {\it Theory of Equations} (McGraw-Hill, NY 1948), 264.

\bibitem{kf}
P. W. Kasteleyn and C. M. Fortuin, J. Phys. Soc. Jpn. (Suppl.) {\bf 26} (1969)
11. 

\bibitem{fk}
C. M. Fortuin and P. W. Kasteleyn, Physica {\bf 57} (1972) 536.

\bibitem{tutte1}
W. T. Tutte, Can. J. Math. {\bf 6} (1954) 80.

\bibitem{tutte2}
W. T. Tutte, J. Combin. Theory {\bf 2} (1967) 301.

\bibitem{tutte3}
W. T. Tutte, ``Chromials'', in Lecture Notes in Math. v. 411
(1974) 243; {\it Graph Theory}, vol. 21 of Encyclopedia of
Mathematics and Applications (Addison-Wesley, Menlo Park, 1984).

\bibitem{hca}
S.-C. Chang and R. Shrock, Physica A {\bf 296} (2001) 183. 

\bibitem{saleur}
H. Saleur, Commun. Math. Phys. {\bf 132} (1990) 657; 
Nucl. Phys. B {\bf 360} (1991) 219.

\bibitem{cf}
S.-C. Chang and R. Shrock, Physica A {\bf 296} (2001) 131. 

\bibitem{sl}
N. J. A. Sloane, {\it The On-Line Encyclopedia of Integer Sequences}, 
\\ http://www.research.att.com/$\sim$njas/sequences/ \ .

\bibitem{Flajolet99}
P. Flajolet and M. Noy, Disc. Math. {\bf 204} (1999) 203.

\bibitem{tk}
N. L. Biggs and R. Shrock, J. Phys. A (Letts) {\bf 32}, L489 (1999).

\bibitem{tor4}
S.-C. Chang and R. Shrock, Physica A {\bf 292} (2001) 307.

\bibitem{rpstanley}
R. P. Stanley, {\it Enumerative Combinatorics}, Vol. 2 (Cambridge University 
Press, Cambridge, 1999).

\bibitem{bernhart99}
F. R. Bernhart, Disc. Math. {\bf 204} (1999) 73.

\bibitem{s5}
S.-C. Chang and R. Shrock, Physica A {\bf 316} (2002) 335.

\bibitem{boll}
B. Bollob\'as, {\it Modern Graph Theory} (Springer, New York, 1998).

\bibitem{t}
S.-C. Chang and R. Shrock, Ann. Phys. {\bf 290}, 124 (2001), cond-mat/0004129.

\bibitem{wcy}
R. Shrock and S.-H. Tsai, Physica A {\bf 275} (2000) 429.

\bibitem{ta}
S.-C. Chang and R. Shrock, Physica A {\bf 286} (2001) 189. 

\bibitem{jz}
S.-C. Chang and R. Shrock,  Physica A {\bf 301} (2001) 196.
            
\bibitem{dg}
S.-C. Chang and R. Shrock, Physica A {\bf 301} (2001) 301. 

\bibitem{sdg}
S.-C. Chang and R. Shrock, Phys. Rev. E {\bf 64} (2001) 066116. 

\bibitem{ka}
S.-C. Chang and R. Shrock, Int. J. Mod. Phys. B {\bf 15} (2001) 443. 

\bibitem{biggs02}
N. L. Biggs, Linear Algebra Appl. {\bf 356} (2002) 3.

\bibitem{ks}
H. Kluepfel and R. Shrock, YITP-99-32; H. Kluepfel, Stony Brook
thesis (July, 1999).

\bibitem{ck}
S.-Y. Kim and R. Creswick, Physical Review E {\bf 63} (2001) 066107. 

\bibitem{f}
S.-C. Chang and R. Shrock, J. Stat. Phys., {\bf 112} (2003) 815.

\end{thebibliography}
\end{document}